\documentclass[a4paper,11pt]{article}
\pdfoutput=1 

\usepackage{jheppub} 

\usepackage[T1]{fontenc} 

\usepackage[usenames,table]{xcolor}
\usepackage{mathrsfs}
\usepackage{tensor}

\def\cD{{\cal D}}

\def\cF{{\cal F}}

\def\cK{{\cal K}}

\def\cM{{\cal M}}

\def\cO{{\cal O}}

\def\cQ{{\cal Q}}

\def\cU{{\cal U}}

\def\a{\alpha} 
\def\b{\beta}  
\def\g{\gamma} 

\def\d{\delta} 

\def\e{\epsilon}

\def\l{\lambda}

\def\m{\mu}
\def\n{\nu}
\def\x{\xi}

\def\vf{\varphi}
\def\O{\Omega}

\def\be{\begin{equation}}
\def\ee{\end{equation}}

\def\vf{\varphi}
\def\pr{\partial}
\def\Ddot{\cD\cdot}
\def\vx{\mathbf{\hat{x}}}
\def\nn{\nonumber}
\def\scri{\mathscr{I}}

\preprint{NORDITA 2020-103}

\title{\bf On asymptotic symmetries in higher dimensions\\ for any spin}

\author[a,1]{Andrea~Campoleoni,\note{Research Associate of the Fund for Scientific Research -- FNRS, Belgium.}}
\author[b,c]{Dario~Francia,}
\author[d,e]{and Carlo~Heissenberg}

\affiliation[a]{Service de Physique de l'Univers, Champs et Gravitation, Universit\'e de Mons, 20 place du Parc, 7000 Mons, 
Belgium}
\affiliation[b]{Centro Studi e Ricerche E.~Fermi,
Piazza del Viminale 1, 00184 Roma, Italy}
\affiliation[c]{Roma Tre University and INFN,
Via della Vasca Navale 84, 00146 Roma, Italy}
\affiliation[d]{Nordita, Stockholm University and KTH Royal Institute of Technology,
Roslagstullsbacken 23, 10691 Stockholm, Sweden}
\affiliation[e]{Department of Physics and Astronomy, Uppsala University, 75108 Uppsala, Sweden}

\emailAdd{andrea.campoleoni@umons.ac.be}
\emailAdd{dario.francia@cref.it}
\emailAdd{carlo.heissenberg@su.se}

\abstract{We investigate asymptotic symmetries in flat backgrounds of dimension higher than or equal to four. For spin two we provide the counterpart of the extended BMS transformations found by Campiglia and Laddha in four-dimensional Minkowski space. We then identify higher-spin supertranslations and generalised superrotations in any dimension. These symmetries are in one-to-one correspondence with spin-$s$ partially-massless representations on the celestial sphere, with supertranslations corresponding in particular to the representations with maximal depth. We discuss the definition of the corresponding asymptotic charges and we exploit the supertranslational ones in order to prove the link  with Weinberg's soft theorem in even dimensions.}

\begin{document} 
\maketitle

\section{Introduction}

 In this work we construct higher-spin supertranslations and generalised superrotations at null infinity, in flat spacetimes of any dimension $D \geq 4$. 
We thus extend the results of \cite{ACD1}, where higher-spin supertranslations and superrotations have been identified in four dimensions, and of \cite{ACD2}, where global higher-spin symmetries have been studied in any $D > 4$.
  
 Following the seminal works  \cite{BMS, Sachs_Waves, Sachs_Symmetries}, the asymptotic symmetry group of four-dimensional asymptotically flat gravity, and later of spin-one gauge theories \cite{Strominger_YM, Barnich_YM, Strominger_QED1, Campiglia_QED}, was long identified as comprising those transformations of the gauge potentials that preserve the falloffs typical of radiation, where the norm of the corresponding fields scales to leading order as $r^{-1}$ in retarded Bondi coordinates. (See \cite{strominger_rev} for a review.) However, in striking contrast with the four-dimensional case, imposing the same requirement in higher-dimensional gravity, where radiation scales asymptotically as $r^{1-\frac{D}{2}}$, effectively selects only the (global) transformations of the Poincar\'e group within the full group of diffeomorphisms, thus apparently preventing BMS$_{D>4}$ to be identified as a physically sensible asymptotic group \cite{BMS_d, Tanabe-Kinoshita-Shiromizu}. The absence of gravitational memory effects to radiative order beyond $D=4$ \cite{BMS-memory}, moreover, provided further support to the idea that $D=4$ was to be regarded as possessing a special status as for what concerns the asymptotic structure of asymptotically flat spacetimes. In the same fashion, in theories of photons or gluons whose associated potentials decay at null infinity as fast as $r^{1-\frac{D}{2}}$ only global $U(1)$ or $SU(N)$ transformations are kept asymptotically. Superrotations, in their turn, were originally identified as the infinite-dimensional family of vector fields providing local solutions to the conformal Killing  equation on the two-dimensional celestial sphere \cite{Barnich_Revisited, Barnich_BMS/CFT}. In this sense, their very existence appeared to be somewhat specific of four-dimensional Minkowski space.
 
A different view was advocated for flat spaces in \cite{Strominger_QED2, Strominger1, Mao:2017wvx, campiglia-scalar-anyD, Strominger2, Afshar:2018apx, Aggarwal:2018ilg, QED_oddD, YM_oddD, capone-taylor, ACD4, Heissenberg:2019fbn, capone_lec, Esmaeili:2020eua}. The interpretation of Weinberg's soft theorems as Ward identities of asymptotic symmetries in $D=4$ rather naturally called for a similar correspondence in higher dimensions, thus suggesting the existence of relevant symmetries beyond the global ones. This picture eventually found two different incarnations. Supertranslations were first recovered in any $D$ by weakening the falloffs of the fields so as to match those of the four-dimensional case \cite{Strominger_QED2, Strominger1, QED_oddD, YM_oddD, capone-taylor}. Memory effects, in their turn, were better identified as due to the leading components of the stationary solutions of the  field equations, whose typical Coulombic scale of $\cO(r^{3-D})$  is subleading with respect to the radiation falloffs in any $D > 4$. While in agreement with the observed absence of memory effects to leading-order, the identification of such higher-dimensional, subleading, memory effects 
also led to consider another class of residual gauge symmetries akin to supertranslations \cite{Strominger2,ACD4} (see also \cite{Mao:2017wvx,campiglia-scalar-anyD}). 
For spin-one gauge theories the presence of angle-dependent asymptotic symmetries in higher dimensions was also confirmed by an analysis at space-like infinity \cite{QED_anyD_spatial,QED_anyD_Hamiltonian}. 

Similarly, the idea that additional asymptotic symmetries, other than supertranslations, could be held responsible for subleading soft graviton theorems \cite{superrotations_soft, sub-soft_anyD} led to identify a different extension of the BMS group in four dimensions as the semidirect product of supertranslations and Diff$(S^2)$ \cite{superrotations, superrotations2}, differently from the original proposal of \cite{superrotations_qg} that would link the subleading soft amplitudes to the Ward identities of the superrotations of \cite{Barnich_Revisited, Barnich_BMS/CFT}. (See also \cite{CM_notsoft} for yet an alternative derivation of subleading soft graviton theorems.) What is relevant to our purposes is that  the four-dimensional construction of \cite{superrotations}, contrary to that of \cite{superrotations_qg, superrotations_soft}, is amenable to be pursued in any $D$ \cite{Avery:2015gxa, superrotations_anyD}. 
 
In the following we apply similar considerations both to low ($s = 1,2$) and to higher-spin ($s > 2$) gauge theories in $D\ge 4$. In \cite{ACD2, ACD3} we showed that, if the asymptotic behaviour typical of radiation is chosen as the leading falloff in $D>4$, the corresponding asymptotic group only comprises the solutions to the global Killing equations.\footnote{Similar conclusions have been drawn for higher-spin fields in Anti de Sitter spacetimes in \cite{HS-AdS-1, HS-AdS-2}.} By contrast, here we begin by imposing in any dimension the same falloffs as those allowing (higher-spin) supertranslations in $D=4$ \cite{ACD1}, i.e.\ we consider fields whose norm scales asymptotically as $r^{-1}$ for any $D \geq 4$. In our Bondi-like gauge \eqref{bondig}, this choice naturally leads to asymptotic symmetries depending on an arbitrary function on the celestial sphere, which we identify as higher-spin counterparts of BMS supertranslations.\footnote{The authors of \cite{Grumiller_softhair} identify  operators performing spin-dependent supertranslations in any $D$ in the analysis of the near-horizon symmetries of a black-hole background, although in the (putative) absence of higher-spin fields. It is conceivable that the chosen class of spin-dependent boundary conditions effectively subsume the presence of higher-spin fields in the corresponding thermal bath.}  In addition, we show that, on shell, all overleading configurations above the falloffs typical of radiation must be pure gauge and then, following \cite{ACD4}, we propose a prescription to associate finite surface charges to higher-spin supertranslations. These results suggest to interpret the additional overleading terms as new \emph{global} degrees of freedom. We complete our analysis of higher-spin supertranslations by showing that Weinberg's factorisation theorems for soft particles of any spin \cite{Weinberg_64,Weinberg_65} can be recovered as Ward identities for these asymptotic symmetries, thus extending to any even space-time dimension the results of \cite{ACD1}. 

We then compute the full set of residual symmetries of the Bondi-like gauge, without any prior assumption on the allowed decay rates of the fields. In this way we discover other classes of infinite-dimensional symmetries that depend on arbitrary traceless tensors on the celestial sphere of rank $1,2 \ldots, s-1$ and, for $s=2$, reduce to the superrotations of \cite{superrotations, superrotations_anyD}. Their scaling with $r$ gets more and more relevant, so that if one wishes to keep all of them the norm of the fields should actually blow up as fast as $r^{s-2}$. However, field configurations that are overleading with respect to radiation can be shown to be anyway pure gauge on shell and thus, for instance, they won't affect the decay rate of the higher-spin Weyl tensors that can be kept to be those typical of radiation. Interestingly, each family of asymptotic symmetries appears to be in one-to-one correspondence with partially massless representations on the celestial sphere \cite{PM_Deser, PM_Higuchi, PM_HS, PM_Zinoviev, PM_AdS/CFT}, identified via the kinetic operators ruling the dynamics of suitable overleading components of the asymptotic field.  Similarly to what happens for gravity in $D=4$, the generalised superrotation charges diverge in the limit $r\to\infty$ and should be properly regularised, in the spirit of \cite{Lambda-BMS-1, Lambda-BMS-2, Lambda-BMS-3}. Here we do not address this issue in its full generality and we limit ourselves to discuss the finiteness of superrotation charges when evaluated on special classes of solutions. In this fashion, for $s=2$, we are at least able to  make partial contact with the charges employed in \cite{superrotations_anyD} to relate the subleading soft graviton theorem and the superrotation Ward identities. 

Higher-spin gauge theories have long been supposed to rule the high-energy limit of string theory and to provide a symmetric phase of the latter, in a regime where the string tension may be taken as negligible \cite{Sagnotti_review}. Whereas the actual import of this tantalising conjecture will remain elusive as long as a concrete mechanism for implementing higher-spin gauge symmetry breaking is not found, still one may hope to highlight glimpses of such hypothetical symmetric phase in possible remnants of higher-spin asymptotic symmetries in string scattering amplitudes. This possibility provides one of the main motivations for the identification of the proper higher-spin asymptotic symmetry group in higher dimensions.

The paper is organised as follows: in Section~\ref{hspst} we focus on supertranslations, presenting the relevant boundary conditions together with our prescription to associate finite surface charges to them. The latter are then used to derive Weinberg's soft theorems for any spin. Some of the relevant results on the structure of asymptotic symmetries are actually proven in Section~\ref{hspsr}, where the scope of our analysis widens  to include higher-spin superrotations. More technical details can be found in the appendices.

\section{Higher-spin supertranslations and Weinberg's soft theorem} \label{hspst}

We consider free gauge fields of spin $s$ on Minkowski spacetime, obeying the Fronsdal equations in the Bondi-like gauge introduced in \cite{ACD1,ACD2}:\footnote{For $s>1$, the Bondi-like gauge \eqref{bondig} is to be interpreted as an on-shell gauge fixing. Indeed, it fixes a number of conditions larger than the number of independent components of the gauge parameter.}
\begin{equation} \label{bondig}
	\varphi_{r\mu_{s-1}}=0=\gamma^{ij}\varphi_{ij\mu_{s-2}}\,.
\end{equation}
Assuming the asymptotic expansion in retarded Bondi coordinates
\be \label{exp-spin-s}
\vf_{u_{s-k}i_k}(r,u,\vx) = \sum_{n} r^{-n}\, U_{i_k}{}^{\!\!(k,n)}(u,\vx) \, ,
\ee
we investigate the asymptotic structure of the gauge symmetries of the form
\be \label{gauge_fronsdal}
\d \vf_{\m_s} = \nabla_{\!\m} \e_{\m_{s-1}} \quad \textrm{with} \quad g^{\a\b} \e_{\a\b\m_{s-3}} = 0
\ee
preserving \eqref{bondig}. In \cite{ACD2} it was shown that, assuming  falloffs not weaker than 
those typical of radiation, i.e.\ $\vf_{u_{s-k}i_k} \vf_{u_{s-k}}{}^{i_k} = {\cal{O}} (r^{2-D})$ or subleading, the resulting asymptotic symmetries  for any spin in $D>4$ comprise only the global solutions to the Killing tensor equations, with no infinite-dimensional enhancement. 

For low spins,  however,  the latter can be recovered upon assuming weaker fall-off conditions that, in the radial gauge for $s=1$ or in the Bondi gauge for $s=2$, essentially amount to accepting asymptotic falloffs as weak as ${\cal{O}} (r^{-1})$ in any $D$ \cite{Strominger_QED2,Strominger1}. Whereas the appropriate choice of falloffs is in itself a gauge-dependent issue,\footnote{See  \cite{CD, Two_form_spatial} for some comments on this point. For Maxwell fields in the Lorenz gauge, for instance, in order to identify an infinite-dimensional asymptotic group it is not sufficient to assume falloffs as weak as ${\cal{O}} (r^{-1})$ and additional  terms  proportional to $\log r$ are needed in any $D\ge4$ \cite{ACD4}.} at the physical level what matters is how to interpret these additional, low-decaying, configurations from the perspective of observables. In \cite{ACD4} we argued that, for $s=1$, no physical inconsistencies arise in considering such weaker falloffs (of the strength needed in the given gauge) as long as all the overleading contributions above the $D-$dimensional radiation behaviour are (large) pure-gauge configurations.
In the following, we shall adopt the same guiding principle. 
In this fashion, certainly no issues can arise for all  gauge-invariant quantities, like the flux of energy per unit retarded time carried by the electromagnetic field or quantities depending on the linearised Weyl tensor for spin two and higher. Nevertheless, the presence of these overleading field components may be source of subtleties in general, as the definition of superrotation charges to be discussed in Section \ref{hspsr} testifies.

With this proviso, in this section we take the same attitude for any spin: we assume overall falloffs as weak as $\vf_{u_{s-k}i_k} \vf_{u_{s-k}}{}^{i_k} = {\cal{O}} (r^{-2})$ for any $s$ in any $D$, we then identify the $u-$independent residual symmetries preserving \eqref{bondig} and we argue that above the radiation order only pure-gauge configurations survive on shell,  while leaving to the next section a detailed derivation of these results. 
Following \cite{ACD1},  we identify such symmetries as higher-spin counterparts of BMS supertranslations.
Explicitly, upon imposing
\be \label{falloff_st}
\vf_{u_{s-k}i_k} = \cO(r^{k-1})
\ee
the residual parameters of the Bondi-like gauge are indeed expressed in terms of an arbitrary function $T(\mathbf{\hat{x}})$ on the celestial sphere. In particular, one first obtains
\begin{equation}\label{synthST}
	\epsilon^{u_{s-k-1}i_k} = r^{-k}\,\frac{(-1)^{s-k-1}(s-k-1)!}{k!(s-1)!}\, \cD^i \cdots \cD^i T(\mathbf{\hat{x}}) + \g^{ii} \mathbb D^{i_{k-2}} T(\mathbf{\hat{x}}) \,,
\end{equation}
where the $\mathbb D_{i_l}$ are suitable rank$-l$ differential operators.  For instance, for $s=3$, one has
\begin{equation} \label{example_3}
        \epsilon^{uu}=T(\mathbf{\hat{x}})\,, \qquad
		\epsilon^{ui}=-\frac{1}{r}\,\pr^i T(\mathbf {\hat x})\,, \qquad
		\epsilon^{ij}=\frac{1}{2r^2}\left[\mathcal{D}^i\mathcal{D}^j-\frac{1}{D}\,\gamma^{ij}(\Delta-2)\right]T(\mathbf {\hat x})\,.
\end{equation}
As discussed in Section~\ref{sec:symm_bondi} and in Appendix~\ref{app:symm}, one can then express the other components of the gauge parameter in terms of those displayed above. Looking at $u$--independent residual symmetries allowed us to focus on supertranslations; removing this assumption while keeping the falloffs \eqref{falloff_st} one finds in addition only the global symmetries discussed in \cite{ACD2}.

We now show how to associate finite surface charges to the symmetries \eqref{synthST}, to be used in the derivation of Weinberg's soft theorems \cite{Weinberg_64,Weinberg_65}. In the Bondi-like gauge \eqref{bondig}, the surface charge at null infinity associated to a gauge transformation is \cite{ACD2}\footnote{The charge defined in \eqref{surface-charge} is equal to $-(s-1)!$ times the charge appearing in Appendix A of \cite{ACD2}.}
\begin{align}
\cQ(u) &  = \lim_{r\to\infty} r^{D-3}\, \sum_{k=0}^{s-1} \binom{s-1}{k}\! \oint d\Omega_{D-2} \bigg\{ (s-k-2)\, \e^{u_{s-k-1}i_k} \left( r\pr_r + D-2 \right) \vf_{u_{s-k}i_k} \nn \\
& + \vf_{u_{s-k}i_k} \left( r\pr_r + D + 2k - 2 \right) \e^{u_{s-k-1}i_k} - \frac{s-k-1}{r}\, \e^{u_{s-k-1}i_k} \Ddot \vf_{u_{s-k-1}i_k} \bigg\} \, ,\label{surface-charge}
\end{align}
which, for $D>4$, naively diverges as $r^{D-4}$ if one evaluates it for the symmetries \eqref{synthST} on field configurations decaying at null infinity as \eqref{falloff_st}. On the other hand, as discussed in Section~\ref{sec:eoms}, the equations of motion imply that asymptotically all contributions above those of a wave solution be pure gauge. On shell one has indeed
\be \label{on-shell_ST}
\vf_{u_{s-k}i_k} = r^{k-1}\, \frac{k(D+k-5)!}{s(D+s-5)!}\, (\Ddot)^{s-k} C_{i_k}{}^{\!\!(1-s)}(\vx) + \cO\!\left(r^{k+1-\frac{D}{2}}\right) \,,
\ee
with the rank--$s$ tensor $C^{(1-s)}$ given by
\be
C_{i_s}{}^{\!\!(1-s)}(\vx) = [(s-1)!]^{-2}\, \cD_i \cdots \cD_i\, \tilde{T}(\mathbf{\hat{x}}) + \cdots \, ,
\ee
where $\tilde T (\mathbf{\hat x})$ is an arbitrary function and the omitted terms implement the traceless projection of the symmetrised gradients, as required by the constraints \eqref{bondig}. Substituting \eqref{synthST} and \eqref{on-shell_ST} in the expression for the surface charge one obtains
\begin{align}
& (-1)^{s-1} \cQ_T(u)  \nn \\
& = \lim_{r\to\infty} r^{D-3} \oint d\Omega_{D-2} \sum_{k=0}^{s-1} \frac{r^{-k}}{k!}\, T\, \Big[ (s-k-2)\,r\pr_r + (s-k-1)(D-k-2) \Big] (\Ddot)^k \vf_{u_{s-k}} \nn \\
& = \lim_{r\to\infty} r^{D-4} \left( \sum_{k=1}^{s-1} \a_k \right) \oint d\Omega_{D-2}\, T\,(\Ddot)^s C^{(1-s)} + \cO(r^{\frac{D-4}{2}}) \, , \label{div1}
\end{align}
with
\be \label{Cs_st}
\a_k = \frac{(D+k-5)!\left[ (k-1)(s-k-2) + (s-k-1)(D-k-2) \right]}{(k-1)!(D+s-5)!} \, .
\ee
Two types of divergences thus arise if one computes the surface charge by first integrating in \eqref{surface-charge} over a sphere at a given retarded time $u$ and radius $r$ and then taking the limit $r \to \infty$.
However, the divergence $\mathcal O(r^{D-4})$ induced by the overleading, pure-gauge contributions actually vanishes because $\sum_{k=1}^{s-1}\a_k = 0$ for any $s$.  The remaining divergence $\mathcal O(r^{\frac{D-4}{2}})$ is related to the presence of radiation: if one assumes that in a neighbourhood of $\mathscr I^+_-$, say for $u < u_0$, there is no radiation and the fields attain a stationary configuration, then the surface charge is finite. A finite charge $\cQ_{T}(u)$ can then be defined for all values of $u$ as the evolution of $\cQ_{T}(-\infty)$ under the equations of motion \cite{ACD4}.\footnote{See also \cite{Freidel:2019ohg} for an alternative procedure to define finite charges in $D>4$ for angle-dependent asymptotic symmetries in Maxwell's electrodynamics and \cite{Aggarwal:2018ilg} for a discussion of supertranslation charges in higher-dimensional gravity.}
 
In order to compute the supertranslation charges, we thus focus on field configurations with the falloffs typical of a stationary solution. Generalising the characterisation of stationary solutions for fields of spin $s \leq 2$ of \cite{memory-anyD-Wald}, we consider
\begin{equation}\label{synthCoul}
	\varphi_{u_{s-k}i_k} = r^{3-D+k}\, \cU_{i_k}{}^{\!\!(k)}(u,\mathbf{\hat{x}}) + \cdots \, .
\end{equation}
As we shall argue in Appendix~\ref{app:stationary}, this choice is tantamount to evaluating the charges on solutions that satisfy $\pr_u U^{(k,n)} = 0$ in the far past of $\scri^+$ for $n \leq D-k-3$. Moreover, in the absence of massless sources, on shell, the rank--$k$ tensors $\cU^{(k)}$ satisfy 
\be\label{eom_div}
(\Ddot)^{k} \cU^{(k)} = 0 \quad \textrm{for} \quad 1 \leq k \leq s \,,
\ee 
as can be checked from \eqref{div_U(k,D-k-3)} where $\mathcal U^{(k)}=U^{(k,D-k-3)}$.
Taking \eqref{synthST}, \eqref{synthCoul} and \eqref{eom_div} into account, the  surface charge \eqref{surface-charge} reads\footnote{For the scalar case, the charges considered in \cite{campiglia-scalar-anyD} formally coincide with \eqref{SurfaceST} evaluated for $s=0$.} 
\begin{equation}\label{SurfaceST}
	\mathcal Q_T(u) = (-1)^{s-1}(D+s-4)\oint d\Omega_{D-2}\,T(\mathbf{\hat{x}})\,\cU^{(0)}(u,\mathbf{\hat{x}})\,,
\end{equation}
which is closely analogous to the expression for the spin-$2$ supertranslation charge in terms of the Bondi mass aspect, $\mathcal Q_T\propto\oint  d\Omega_{D-2}\,T\, m_B$. 

To summarise, assuming that the fields be on shell up to the falloffs of stationary solutions and defining the charges according to the prescription of \cite{ACD4}, one obtains finite supertranslation charges for any value of $s$ and in any $D$.
Furthermore, let us note that a pure supertranslation configuration carries away no energy to $\mathscr I^+$ per unit retarded time, as defined via the canonical stress-energy tensor $t_{\alpha\beta}$ stemming from the Fronsdal Lagrangian, which in the in the gauge \eqref{bondig} takes the Maxwell-like form \cite{Gen_connections, maxwell-like}
	\begin{equation}\label{}
		\mathcal L = - \frac{\sqrt{-g}}{2}\left(
		\nabla_{\alpha} \varphi_{\mu_{s}} \nabla^{\alpha} \varphi^{\mu_{s}}-s \nabla \cdot \varphi_{ \mu_{s-1}}  \nabla \cdot \varphi^{\mu_{s-1}}
		\right).
	\end{equation}
Indeed, the canonical stress-energy tensor obtained from this Lagrangian reads
\begin{equation}\label{talphabeta}
	t_{\alpha\beta} = \frac{1}{2}\left( \nabla_{\!\alpha}\, \varphi^{\mu_s} \nabla_{\!\beta}\, \varphi_{\mu_s}- s\, \nabla \cdot \varphi^{\mu_{s-1}} \nabla_{\!\alpha}\,  \varphi_{\beta\mu_{s-1}} \right) + g_{\alpha\beta}\left(\,\cdots\right),
\end{equation}
while the energy flux at a given retarded time $u$ is given by
\begin{equation}\label{}
	\mathcal P(u) = \lim_{r\to\infty} \oint \left( t_{uu}-t_{ur}\right) d\Omega_{D-2}\,.
\end{equation}
In the latter expression, the term of the stress-energy tensor \eqref{talphabeta} proportional to the background metric $g_{\alpha\beta}$ drops out, while the remaining ones involve derivatives with respect to $u$. Pure supertranslations however are $u$-independent, and therefore eventually provide a vanishing contribution.

Let us also rewrite the surface charge evaluated at $\mathscr I^+_-$ in terms of an integral over $\mathscr I^+$ according to
\begin{equation} \label{charge-integral}
	\mathcal Q_T\big |_{\mathscr I^+_-} = 
	\mathcal Q_T\big|_{\mathscr I^+_+}-\int_{-\infty}^{+\infty} \frac{d\mathcal Q_T(u)}{du}\,du\,,
\end{equation}
where the first contribution accounts for the presence of stable massive particles in the theory. In their absence, making use of \eqref{SurfaceST}, one finds
\begin{equation}\label{chargescri}
	\mathcal Q_T\big |_{\mathscr I^+_-} = (-1)^{s} \,(D+s-4)\int_{-\infty}^{+\infty}du \oint d\Omega_{D-2}\, T(\mathbf{\hat x})\,\partial_u\, \cU^{(0)}(u,\mathbf{\hat x})\,.
\end{equation}
We can now connect the charge \eqref{chargescri} to Weinberg's soft theorem in even $D$.
As usual, the strategy is to express the Coulombic contributions  appearing in the charge in terms of the radiative contributions 
making use of the equations of motion, so as to make contact with the free field oscillators naturally contained in the radiation components. 
The soft
theorem can then be retrieved by simplifying the insertions of these 
operators in the corresponding Ward identities so as to highlight the 
factorisation of $S$-matrix elements that takes place in the soft limit

Let us consider the spin-three case first. The equations of motion in the Bondi-like gauge allow one to express the charge \eqref{chargescri} in terms of the spin-three generalisation of the Bondi news tensor via
\begin{equation}\label{EOMn33}
	\partial_u^{\frac{D-4}{2}} \mathcal U^{(0)}
	=
	\frac{\mathscr D \left(\mathcal{D}\cdot\right)^3 C^{\left(\tfrac{{D-8}}{2}\right)}}{(D-1)({D-2})(D-3)}\,,
\end{equation}
where the operator $\mathscr D$ is defined as 
\begin{equation}\label{Dcalligraficol}
	\mathscr D = \prod_{l=\frac{{D}}{2}}^{D-3}\mathscr D_l\,,
	\qquad
	\text{with}\quad 
	\mathscr D_{l}=\frac{\Delta-(l-1)(D-l-2)}{D-2l-2} \,.
\end{equation}
One can therefore rewrite the charge \eqref{chargescri} as follows 
\begin{equation}\label{chargescri333}
\begin{split}
	& \mathcal Q_T\big |_{\mathscr I^+_-} =  -\frac{1}{(D-2)(D-3)} \int_{-\infty}^{+\infty}du \oint d\Omega_{D-2}\,T (\mathbf{\hat x})\,\partial_u^{\frac{6-D}{2}}\mathscr D\, \mathcal{D}^i  \mathcal{D}^j \mathcal{D}^k C_{ijk}^{\left(\tfrac{D-8}{2}\right)}\!(u,\vx) \\[5pt]
	& \hspace{10pt} = \frac{1}{4(D-2)(D-3)} \lim_{\omega\to0^+} \sum_\lambda \oint \frac{d\Omega_{D-2}}{(2\pi)^{\frac{D-2}{2}}}\,T(\mathbf{\hat x}) \mathscr D \mathcal{D}^i \mathcal{D}^j \mathcal{D}^k \epsilon_{ijk}^{(\lambda)}(\mathbf{\hat x})\,\omega a_\lambda(\omega \mathbf{\hat x})+ \textrm{H.c.} \,,
\end{split}
\end{equation}
where in the last equality we inserted the expansion in oscillators of the leading radiation contribution to $\varphi_{ijk}$, 
\begin{equation}\label{}
	C_{ijk}^{\left(\tfrac{D-8}{2}\right)}\!(u,\vx)
	=
	\frac{1}{2(2i\pi)^{\frac{D-2}{2}}}
	\int_0^\infty
	\frac{d\omega}{2\pi} \, \omega^{\frac{D-4}{2}} e^{-i\omega u} \sum_\lambda \epsilon_{ijk}^{(\lambda)}(\mathbf{\hat x})\,\omega a_\lambda(\omega \mathbf{\hat x}) + \textrm{H.c.}\,,
\end{equation}
and we used the relations
\begin{align}\label{}
	(i\partial_u)^{\frac{6-D}{2}}\int_0^\infty \omega^{\frac{D-4}{2}}e^{-i\omega u} f(\omega) d\omega &= \int_0^\infty \omega e^{-i\omega u} f(\omega) d\omega \,,\\[5pt]
	\frac{1}{2\pi} \int_{-\infty}^{+\infty}du\int_{0}^{\infty} \omega e^{-i\omega u} f(\omega)\, d\omega &= \frac{1}{2}\,\lim_{\omega\to 0^+} \left[\omega f(\omega)\right].
\end{align}
The charge \eqref{chargescri333} enters the Ward identity
\begin{equation} \label{Ward3}
	\langle\text{out}| \left(\mathcal Q_{\mathscr I^+_-} S - S \mathcal Q_{\mathscr I^-_+}\right) |\text{in}\rangle = \sum_\ell g^{(3)}_\ell E^2_\ell\, T(\mathbf{\hat x}_\ell) \langle\text{out}| S |\text{in}\rangle\,,
\end{equation}
under the assumption that higher-spin supertranslations are symmetries of a putative scattering matrix involving particles with arbitrary spins. More precisely, we follow the procedure detailed in \cite{Avery-Schwab, ACD1} for connecting the soft portion of the asymptotic charge to the Ward identity \eqref{Ward3}, which avoids the need to explicitly discuss external currents. In order to highlight the relation to Weinberg's soft theorem it is useful to choose a specific form for the function $T(\mathbf{\hat x})$:
\begin{equation} \label{Tspec}
	T_{\mathbf{\hat w}}(\mathbf{\hat x})  = (\mathcal{D}_{\mathbf{\hat w}})_i (\mathcal{D}_{\mathbf{\hat w}})_j (\mathcal{D}_{\mathbf{\hat w}})_k \frac{\epsilon^{(ijk)}_{lmn}(\mathbf{\hat w})(\mathbf{\hat x})^l(\mathbf{\hat x})^m(\mathbf{\hat x})^n}{1-\mathbf{\hat x}\cdot \mathbf{\hat w}}\,,
\end{equation}
where the choice of polarisations is discussed in Appendix \ref{app:polarizations}. Inserting \eqref{Tspec} in \eqref{chargescri333} one finds 
\begin{equation}
	\mathcal Q_{T_{\mathbf{\hat w}}}\big |_{\mathscr I^+_-} = -\frac{1}{2} \lim_{\omega\to0^+} \mathcal{D}_{\mathbf{\hat w}}^i \mathcal{D}_{\mathbf{\hat w}}^j \mathcal{D}_{\mathbf{\hat w}}^k \left[\,\omega a_{ijk} (\omega \mathbf{\hat w})+\omega a_{ijk}^\dagger(\omega \mathbf{\hat w})\,\right].
\end{equation}
Substituting this relation into the Ward identity \eqref{Ward3} then yields the 3-divergence of Weinberg's theorem, 
\begin{equation} \label{soft}
	\lim_{\omega\to0^+} \langle\text{out}| \omega a^{ijk} (\omega\mathbf {\hat w}) S |\text{in}\rangle 
	=  -\sum_\ell g_\ell^{(3)} E_\ell^2  \,
	\frac{\epsilon^{(ijk)}_{lmn}(\mathbf{\hat w})(\mathbf{\hat x}_\ell)^l (\mathbf{\hat x}_\ell)^m (\mathbf{\hat x}_\ell)^n}{1-\mathbf{\hat w}\cdot \mathbf{\hat x}_\ell} \,
	\langle\text{out}| S |\text{in}\rangle\,.
\end{equation}
This argument holds for any values of the couplings $g_\ell^{(3)}$, thus showing that the relation between the Ward identity and the soft theorem is actually universal and does not rely on the actual possible dynamical incarnations of the theory itself.

The proof extends verbatim to the spin-$s$ case. One starts with the charge
\begin{equation}\label{chargescris}
\begin{split}
	\mathcal Q_T\big |_{\mathscr I^+_-} &= (-1)^{s}(D-4+s) 
	\int_{-\infty}^{+\infty}du \oint d\Omega_{D-2}\,T\,\partial_u \cU^{(0)} \,,
\end{split}
\end{equation}
and makes repeated use of the equations of motion, using in particular
\begin{equation}\label{}
	\pr_u^{\frac{D-4}{2}} \cU^{(0)} = \frac{(D-4)!}{(D+s-4)!}\, \mathscr D \, (\Ddot)^s C^{(\frac{D-2s-2}{2})}\,,
\end{equation} 
to put it in the form 
\be
\begin{split}
	\mathcal Q_T\big |_{\mathscr I^+_-} & = \frac{(-1)^{s}(D-4)!}{(D+s-5)!} \int_{-\infty}^{+\infty}du \oint d\Omega_{D-2}\,T\,\partial_u^{2-\frac{{D-2}}{2}}\mathscr D (\mathcal{D}\cdot)^s C^{\left(\frac{D-2s-2}{2}\right)} \\[5pt]
	& = \frac{(-1)^{s-1}(D-4)!}{4(D+s-5)!} \lim_{\omega\to0^+} \sum_\lambda \oint \frac{d\Omega_{D-2}}{(2\pi)^{\frac{D-2}{2}}}\,T \mathscr D (\mathcal{D}\cdot)^s \epsilon^{(\lambda)}(\mathbf{\hat x})\omega a_\lambda(\omega \mathbf{\hat x})+\textrm{H.c.}\,, \label{chargescrisss}
\end{split}
\ee
where in the last equality we substituted the asymptotic limit of the free field near $\mathscr I^+$ while the operator $\mathscr D$ is defined as in \eqref{Dcalligraficol}. In order to connect the Ward identity of higher-spin supertranslations
to the soft theorem it is useful once again to make use of a specific form of $T(\mathbf{\hat x})$,
\begin{equation} \label{T_s}
	T_{\mathbf{\hat w}}(\mathbf{\hat x})  = (\mathcal{D}_{\mathbf{\hat w}})_{i_s} \frac{\epsilon^{(i_s)}_{j_s}(\mathbf{\hat w})(\mathbf{\hat x})^{j_s}}{1-\mathbf{\hat x}\cdot \mathbf{\hat w}}\,, 
\end{equation}
in terms of which the charge reads
\begin{equation}
	\mathcal Q_{T_{\mathbf{\hat w}}}\big |_{\mathscr I^+_-} = -\frac{1}{2} \lim_{\omega\to0^+} \mathcal{D}_{\mathbf{\hat w}}^{i_s} \left[\, \omega a_{i_s}\! (\omega \mathbf{\hat w})+\omega a_{i_s}^\dagger\!(\omega \mathbf{\hat w}) \,\right].
\end{equation}
Substituting this relation into the spin-$s$ version of the Ward identity \eqref{Ward3} then yields the $s$--divergence of Weinberg's theorem. The reverse implication, on the other hand, namely that Weinberg's theorem yields the Ward identity \eqref{Ward3} as well as its spin-$s$ counterpart, is of less relevance in the context of higher spins given that Weinberg's result also implies the vanishing of the soft couplings for $s>2$. 

\section{Higher-spin superrotations} \label{hspsr}

In this section we classify all residual symmetries of the Bondi-like gauge \eqref{bondig} and we show that they comprise, in any dimension and for any value of the spin, suitable generalisations of the superrotations introduced for $s=2$ and $D=4$ in \cite{superrotations}. In particular, within the limits of our linearised analysis, for $s=2$ we find extended BMS symmetries comprising both supertranslations and Diff($S^{D-2}$) transformations as in \cite{superrotations_anyD}. For arbitrary values of the spin we find instead asymptotic symmetries generated by a set of traceless tensors on the celestial sphere of rank $0,1,\ldots,s-1$, that turn out to be in one-to-one correspondence with the partially massless representations of spin $s$, with supertranslations corresponding in particular to the representations with maximal depth.
To keep all such residual symmetries of the Bondi-like gauge, the non-vanishing components of the fields must scale as
\be \label{falloffs_superrotations}
\vf_{u_{s-k}i_k} = \cO(r^{s+k-2}) \, , 
\ee
although, eventually, only pure-gauge contributions are allowed on shell above the order typical of a radiative solution, $\vf_{u_{s-k}i_k} =\cO(r^{k+1-D/2})$.  Still, the definition of surface charges for (higher-spin) superrotations  entails a number of subtleties that here we are able to face only to a partial extent and that require further investigations. 

\subsection{Symmetries of the Bondi-like gauge} \label{sec:symm_bondi}

We begin by identifying the residual symmetries allowed by the Bondi-like gauge \eqref{bondig}, without any further specifications on the falloffs of the components $\vf_{u_{s-k}i_k}$. To this end, it is convenient to split the components of the gauge parameter in two groups: those without any index $u$, that we denote by $\e_{i_k}{}^{\!\!(k)} \equiv \e_{r_{s-k-1}i_k}$, and the rest. Notice that not all components are independent because the gauge parameter is traceless: here we chose to express those with at least one index $r$ and one index $u$ in terms of the others.

The elements of the first group are constrained by
\be \label{variation-radial-s}
\d \vf_{r_{s-k}} = \frac{1}{r} \left\{ (s-k) \left( r\pr_r - 2k \right) \e^{(k)} - \g\, \e^{(k)\prime} \right\} + \cD \e^{(k-1)} + r\,\g\, \e^{(k-2)} = 0 \, ,
\ee
where a prime denotes a contraction with $\g^{ij}$ and where we omitted all sets of symmetrised angular indices.  These equations are solved by
\be \label{ansatz-parameter}
\e^{(k)}(r,u,\vx) = r^{2k} \rho^{(k)}(u,\vx) + \sum_{l\,=\,k}^{2k-1} r^l \e^{(k,l)}(u,\vx) \, ,
\ee
where, at this stage, $\rho^{(k)}(u,\vx)$ is an arbitrary traceless tensor because $r^{2k}$ belongs to the kernel of $(r\pr_r-2k)$. It is however bound to be traceless because of Fronsdal's trace constraint. The $\e^{(k,l)}$ are instead determined recursively (and algebraically) in terms of the $\rho^{(l)}$ with $l < k$. The precise form of the tensors $\e^{(k,l)}$ is not relevant for the ensuing considerations; we thus refer to Appendix~\ref{app:symm} for more details.

One can express the remaining components $\e_{u_{s-k-1}i_k}$ in terms of the $\rho^{(k)}$ by imposing that all traces of the fields be gauge invariant, i.e.\ $
\g^{mn} \d \vf_{u_{s-k}i_{k-2}mn} = 0$. Imposing $\d \vf_{r_{s-k}u_li_k} = 0$ for $k+l<s$ leads instead to a constraint on the free tensors in \eqref{ansatz-parameter}:
\be \label{diff_u}
\pr_u \rho^{(k)} + \frac{s-k-1}{D+s+k-4}\, \Ddot \rho^{(k+1)} = 0
\ee
for any $k<s-1$ (see Appendix~\ref{app:symm}).

For a field of spin $s$, we thus obtain residual symmetries parameterised by the $s-1$ traceless tensors on the celestial sphere $\rho^{(0)}$, $\rho_i^{(1)}$, \ldots, $\rho_{i_{s-1}}^{(s-1)}$, where the tensor of highest rank still admits an arbitrary dependence on $u$. As we shall see in the next subsection, one can eliminate the $u$--dependence by demanding that $\vf_{i_s}$ falloffs as fast as the $\d \vf_{i_s}$ induced by \eqref{ansatz-parameter} and imposing the equations of motion above the radiation order. Under these assumptions, one obtains the falloffs \eqref{falloffs_superrotations}, while the differential equation \eqref{diff_u} holds for any value of $k$, so that $\pr_u \rho^{(s-1)} = 0$.

When $s = 2$, the residual symmetries of the Bondi gauge $h_{rr} = h_{ru}= h_{ri} = \g^{ij} h_{ij} = 0$ are generated by
\be \label{symm2}
\e_r = f \, , \qquad
\e_i = r^2 v_i + r\, \pr_i f \, , \qquad
\e_u = \e_r + \frac{r^{-1}}{D-2}\, \Ddot \e \, ,
\ee
with the constraint
\be
\pr_u f + \frac{1}{D-2}\, \Ddot v = 0 \, .
\ee
Imposing the falloffs \eqref{falloffs_superrotations}, that is $h_{ij} = \cO(r^{2})$, $h_{ui} = \cO(r)$ and $h_{uu} = \cO(1)$, one obtains the additional condition
\be
\pr_u v_i = 0 \quad \Rightarrow \quad f(u,\vx) = T(\vx) - \frac{u}{D-2}\, \Ddot v(\vx) \, .
\ee
For any value of the space-time dimension, we thus recovered the supertranslations discussed in the previous section, together with a transformation generated by a free vector on the celestial sphere. To leading order, the latter acts on $h_{ij}$ as the traceless projection of a linearised diffeomorphism, 
\be
\d h_{ij} = r^2 \left( \cD_{(i} v_{j)} - \frac{2}{D-2}\, \g_{ij} \Ddot v \right) + \cO(r) \, .
\ee   
In a full, non-linear theory this transformation corresponds to the superrotations of \cite{superrotations,superrotations_anyD} (see also \cite{Adami:2020ugu} for a related discussion).

This pattern continues for arbitrary values of the spin. For instance, for $s=3$ the residual symmetries of the Bondi-like gauge are generated by
\begin{subequations} \label{symm3}
\begin{align}
\e_{rr} & = f \, , \label{e_rr} \\[5pt]
\e_{ri} & = r^2 v_i + \frac{r}{2}\, \pr_i f \, , \label{e_ri} \\
\e_{ij} & = r^4 K_{ij} + r^{3}  \left( \cD_{(i} v_{j)} - \frac{2}{D-1}\, \g_{ij} \Ddot v \right) + \frac{r^2}{2} \left( \cD_i \cD_j - \frac{1}{D}\, \g_{ij} \left( \Delta - 2 \right) \right) f , \label{e_ij}
\end{align}
\end{subequations}
where $K_{ij}$ must be traceless to fulfil the constraint $g^{\m\n}\e_{\m\n} = 0$, while
\be \label{diff_u_3}
\pr_u f + \frac{2}{D-1}\, \Ddot v = 0 \, , \qquad
\pr_u v_i + \frac{1}{D}\, \Ddot K_i = 0 \, .
\ee
Out of the remaining components of the gauge parameter one finds  
	$
		2\e_{ru} = \e_{rr} + r^{-2} \e' 
	$,
while the conditions $\g^{jk} \d \vf_{ijk} = 0$ and $\g^{ij} \d \vf_{uij} = 0$ imply, respectively,
\begin{subequations}
\begin{align}
\e_{ui} & = \e_{ri} + \frac{r^{-1}}{2D} \left( 2\,\Ddot \e_i + \cD_i \e' \right) , \qquad 
\e_{uu} = \e_{ru} + \frac{r^{-1}}{2(D-2)} \left( 2\,\Ddot \e_u + \pr_u \e' \right) . 
\end{align}
\end{subequations}
Imposing the boundary conditions \eqref{falloffs_superrotations} then selects the following solution for \eqref{diff_u_3}:
\begin{subequations} \label{parameters3}
\begin{align}
K_{ij} & = K_{ij}(\vx) \, , \\[5pt]
v_i & = \rho_i(\vx) - \frac{u}{D}\, \Ddot K_i(\vx) \, , \\
f & = T(\vx) - \frac{2u}{D-1}\, \Ddot \rho(\vx) + \frac{u^2}{D(D-1)}\, \Ddot\Ddot K(\vx) \, . 
\end{align}
\end{subequations}
As expected, keeping only the $u$--independent contributions forces $\rho_i=0$ and $K_{ij}=0$ so that one recovers \eqref{example_3}.

In Bondi coordinates, the solutions of the Killing tensor equation $\d \vf_{\m_s} = \nabla_\m \e_{\m_{s-1}} = 0$ take the same form, but the tensors $K_{ij}$, $\rho_i$ and $T$ are bound to satisfy the following additional (traceless) differential constraints, that only leave a finite number of solutions for $D>4$ \cite{ACD2}:
\begin{subequations} \label{diff_constr}
\begin{align}
& \cD_{(i} K_{jk)} - \frac{2}{D}\, \g_{(ij} \Ddot K_{k)} = 0 \, , \label{DK} \\
& \cD_{(i} \cD_{j} \rho_{k)} - \frac{2}{D}\, \g_{(ij} \left[ \left( \Delta + D -3 \right) \rho_{k)} + 2\, \cD_{k)} \Ddot \rho \right] = 0 \, , \label{DDrho}\\
& \cD_{(i} \cD_{j} \cD_{k)} T - \frac{2}{D}\, \g_{(ij} \cD_{k)} \left( 3\,\Delta + 2(D-3) \right) T = 0 \, .
\end{align}
\end{subequations}
With the boundary conditions \eqref{falloffs_superrotations} we thus observe an infinite-dimensional enhancement of all classes of higher-spin symmetries appearing in \eqref{symm3}, but of a different kind compared to the higher-spin superrotations introduced for $D=4$ where, following the spin-2 proposal of \cite{Barnich_BMS/CFT}, we showed that the first two constraints in \eqref{diff_constr} admit locally an infinite-dimensional solution space \cite{ACD1}.

\subsection{Equations of motion above the radiation order} \label{sec:eoms}

We now show that on shell only local pure-gauge field configurations are allowed above the radiation order for fields that admit the asymptotic expansion \eqref{exp-spin-s}.\footnote{For $s=1$ we thus show that the pure-gauge configurations above the usual radiation falloffs that we introduced in \cite{ACD4} exhaust all solutions of the equations of motions. See also \cite{memory-anyD-Wald} for a similar analysis of the equations of motion for $s \leq 2$.} Let us stress that most of the conclusions in this section apply to both even and odd values of the space-time dimension $D$, with the proviso that in the latter case one also has to consider half-integer values of $n$. For simplicity, however, in the following we focus on the case of even $D$, and thus consider $n \in \mathbb{Z}$. See also \cite{capone-taylor} for the corresponding analysis in $D= 5$.

We study the equations of motion above the falloffs typical of radiation discussed in \cite{ACD2}, and in this range matter sources cannot contribute. Furthermore, since the number of angular indices carried by each tensor $U^{(k,n)}$ appearing in the radial expansion \eqref{exp-spin-s} is equal to $k$, from now on we shall omit them altogether.  Introducing the shorthand $C^{(n)} \equiv U^{(s,n)}$, the source-free Fronsdal equations in the Bondi-like gauge imply 
\be \label{U(k,n)}
U^{(k,n)} = \frac{(n+2k-1)(D-n-4)!}{(n+s+k-1)(D-n+s-k-4)!}\, (\Ddot)^{s-k} C^{(n-s+k)}
\ee
for $2-s-k \leq n \leq D-4$, and
\be \label{dCs}
\begin{split}
& (D-2n-2s-2) \pr_u C^{(n)} = \left[ \Delta - (n-1)(D-n-2s-2) - s(D-s-2) \right] C^{(n-1)} \\[3pt]
& \hspace{15pt} - \frac{D+2(s-3)}{(n+2s-2)(D-n-3)} \left( \cD \Ddot C^{(n-1)} - \frac{2}{D+2(s-3)}\, \g\, \Ddot\Ddot C^{(n-1)} \right)
\end{split}
\ee
for $3-2s \leq n \leq D-4$. Out of the specified ranges of $n$, some of the $U^{(k,n)}$ may not be expressed solely in terms of the $C^{(n)}$ and they satisfy differential equations in $u$ similar to \eqref{dCs} (see Appendix~\ref{app:stationary}).

The last equation shows that $C^{\left(\frac{D-2s-2}{2}\right)}(u,\vx)$ is an arbitrary function, corresponding to the ``radiation order''. For $n = \frac{D-2s-2}{2}$ one thus obtains
\be \label{first-inv}
0 = \left[ \Delta - \frac{(D-2s-2)(D-2s-4)}{4} - s(D-s-2) \right] C^{(\frac{D-2s-4}{2})} + \cdots \, ,
\ee
that, on a compact manifold like the celestial sphere, implies $C^{(\frac{D-2s-4}{2})} = 0$. One can reach this conclusion by first eliminating the divergences of the tensor via the divergences of \eqref{first-inv}, and then by noticing that the differential operator $(\Delta - \l)$ entering \eqref{first-inv} is invertible. This is so because the eigenvalues of the Laplacian acting on a traceless and divergenceless tensor of rank $s$ are always negative (see \eqref{eigenvalues}), while $\l > s$.

The previous procedure can be iterated to get
\be \label{zero-Cs}
C^{(n)} = 0 \quad \textrm{for} \quad 1-s < n < \frac{D-2s-2}{2} \, ,
\ee
where the two extrema correspond to the radiation order and to the order at which supertranslations act on the purely angular component $\vf_{i_s}$, respectively. Notice that they coincide when $D=4$ for any value of the spin: in this case supertranslations act at the radiation order, that in the Bondi-like gauge encodes information about the local degrees of freedom of a propagating wave packet \cite{ACD1}. 
To prove \eqref{zero-Cs} it is useful to compute the divergences of \eqref{dCs}:
\be \label{div-dCs}
\begin{split}
&(D-2n-2s-2) \pr_u (\Ddot)^k C^{(n)} = \frac{(n+2s-k-2)(D-n-k-3)}{(n+2s-2)(D-n-3)}\, \times \\
& \hspace{8pt} \times \left[ \Delta - (n+k-1)(D-n-2s+k-2) - (s-k)(D-s+k-2) \right] (\Ddot)^k C^{(n-1)} \\[2pt]
& \hspace{8pt} - \frac{D+2(s-k-3)}{(n+2s-2)(D-n-3)} \left( \cD - \frac{2}{D+2(s-k-3)}\, \g\, \Ddot \right) (\Ddot)^{k+1} C^{(n-1)} .
\end{split}
\ee
For $\frac{D-2s-4}{2} \leq n \leq 3-s$ these equations set to zero recursively all divergences of $C^{(n-1)}$ and eventually the whole tensor itself since all operators in the second line are invertible.
To make this analysis more transparent it is convenient to let $n=D-s-2+\ell$, so that \eqref{div-dCs} takes the form
\begin{align}
	&(2-D-2\ell) \partial_u (\mathcal D\cdot )^k C^{(D-s-2+\ell)} \nn \\
	=&\ \frac{(D-4+s+\ell-k)(s-1-\ell-k)}{(D-4+s+\ell)(s-1-\ell)}
	\left[ \Delta + \ell(\ell+D-3) - (s-k)\right] (\Ddot)^{k} C^{(D-s-3+\ell)} \nn \\
	&- \frac{D+2(s-k-3)}{(D-4+s+\ell)(s-1-\ell)} \left( \cD - \frac{2}{D+2(s-k-3)}\, \g\, \Ddot \right) (\Ddot)^{k+1} C^{(D-s-3+\ell)}\,. \label{div-Cs-ell}
\end{align}
The values of $\ell$ at which the operator appearing in the second line fails to be invertible are
\begin{equation}\label{}
	\ell = 4-D+k-s\,,\qquad \ell =s-1-k\,,
\end{equation}
where the overall coefficient vanishes, or
\begin{equation}\label{}
	\ell=s-k,\, s-k+1,\, s-k+2,\ldots\,,
\end{equation}
as dictated by the eigenvalues of the Laplacian on divergence-free tensors (see \eqref{eigenvalues}).

The iterative procedure that sets the $C^{(n)}$ to zero thus stops at $n=2-s$, since for $k=s$ the overall coefficient in the second line of \eqref{div-dCs} vanishes and one does not obtain any information on $(\Ddot)^s C^{(1-s)}$. 
This result agrees with those of Section~\ref{sec:symm_bondi}: one cannot conclude $C^{(1-s)} = 0$ because under a supertranslation this tensor transforms as
\be\label{memory?}
\d C^{(1-s)} = [(s-1)!]^{-2}\, \cD^s T + \cdots \, ,
\ee 
where the omitted terms implement a traceless projection.\footnote{Let us note that the operator implicitly defined in \eqref{memory?} provides the spin-$s$ counterpart of the differential operator computing the linear memory effect in terms of the supertranslation parameter for spin two in $D=4$, where it indeed acts at the correct Coulombic order \cite{memory}.} This can be easily verified for $s=2$ and $s=3$ by substituting \eqref{symm2} in $\d h_{ij}$ and \eqref{symm3} in $\d \vf_{ijk}$.

This phenomenon extends to all instances of \eqref{dCs} in the range $3-2s \leq n \leq 2-s$. To make this manifest, let us relabel $n \to 2-s-t$: the rhs of \eqref{dCs} then becomes
\be \label{M(s,t)} 
\begin{split}
& \cM^{(s,t)} \equiv \left[ \Delta - (D+s-4)+t(D+t-5) \right] C^{(1-s-t)} \\[3pt]
& \hspace{15pt} - \frac{D+2(s-3)}{(s-t)(D+s+t-5)} \left( \cD \Ddot C^{(1-s-t)} - \frac{2}{D+2(s-3)}\, \g\, \Ddot\Ddot C^{(1-s-t)} \right) ,
\end{split}
\ee
and in the range $0 \leq t \leq s-1$ eq.~\eqref{div-dCs} implies
\be \label{bianchi_k}
(\Ddot)^{s-t} \cM^{(s,t)} = - \frac{D+2(t-3)}{(s-t)(D+s+t-5)} \left( \cD - \frac{2}{D+2(t-3)}\, \g\, \Ddot \right) (\Ddot)^{s-t+1} C^{(1-s-t)} \, .
\ee
The latter can be interpreted as a Bianchi identity for the operator $\cM^{(s,t)}$ and, indeed, it allows one to prove that it is invariant under 
\be \label{gauge_PM}
\d C^{(1-s-t)} = \cD^{s-t} \lambda^{(t)} \quad \textrm{with} \quad \Ddot \lambda^{(t)} = \lambda^{(t)\,\prime} = 0 \, .
\ee
In our context, these transformations can be identified with the portion of the asymptotic symmetries generated by the $u$--independent and divergence-free part of the parameters \eqref{ansatz-parameter}. The other contributions to the residual symmetries of the Bondi-like gauge are reinstated by the sources on the lhs of \eqref{div-dCs}, while their action on the other non-vanishing components of the field, i.e.\ $\d \vf_{u_{s-k}i_k}$, can be recovered from \eqref{U(k,n)} since gauge symmetries map solutions of the eom into other solutions.
For instance, for $s=2$ one obtains
\begin{subequations} \label{spin-2_overleadingC}
\begin{align}
\d C_{ij}{}^{(-2)} & = \cD_{(i} v_{j)} - \frac{2}{D-2}\, \g_{ij} \Ddot v \, , \\
\d C_{ij}{}^{(-1)} & = 2 \left( \cD_i \cD_j - \frac{1}{D-2}\,\gamma_{ij} \Delta \right) f ,
\end{align}
\end{subequations}
and, correspondingly,
\begin{subequations} \label{spin-2_overlaeding}
\begin{align}
\d h_{ui} & = \frac{\Ddot \d C_i{}^{(-1)}}{2(D-3)} = \frac{\cD_i \left( \Delta + D - 2 \right) f}{D-2} \, , \\
\d h_{uu} & = - \frac{\Ddot \Ddot \d C^{(-2)}}{(D-2)(D-3)} = - \frac{2 \left( \Delta + D - 2 \right) \Ddot v}{(D-2)^2} \, .
\end{align}
\end{subequations}

Let us also observe that the differential operator $(\Delta - m^2_{s,t})$ in \eqref{M(s,t)} identifies the mass shell of a partially-massless field of spin $s$ and depth $t$ (see e.g.~\cite{PM_Zinoviev}). Moreover, for $t=s-1$ one recovers in \eqref{M(s,t)} the Maxwell-like kinetic operator for a massless field of spin $s$ propagating on a constant curvature background \cite{maxwell-like}, while for the other values of $t$ one obtains kinetic operators describing more complicated spectra. In particular, for $s=2$ and $t=0$ eq.~\eqref{div-dCs} gives the conformally-invariant equation of motion introduced in \cite{Drew:1980yk}, that does not describe only a partially-massless spin-2 field.
 
We now impose the additional condition that the field components above the order at which asymptotic symmetries act be zero, that is
\begin{equation}\label{0overoverl}
C^{(n)}=0 \quad \textrm{for} \quad n < 2-2s\,,
\end{equation}
or more generally $U^{(k,n)}=0$ for $n < 2-s-k$. This corresponds to the boundary conditions \eqref{falloffs_superrotations}.
Thanks to \eqref{zero-Cs}, under this assumption the only non-vanishing $C^{(n)}$ above the radiation order are those with $2-2s \leq n \leq 1-s$ and we wish to argue that only the pure-gauge configurations that we discussed above satisfy the equations of motion. 

In order to support this statement, let us examine in detail the low-spin examples.
For spin one, the only nontrivial overleading component is $C_i{}^{(0)}$ and it satisfies the free Maxwell equation on the Euclidean sphere
\begin{equation}\label{MaxwellSphere}
\left[\Delta - D + 3\right] C_i{}^{(0)}-\mathcal D_i\mathcal D\cdot C^{(0)}=0\,.
\end{equation}
We can separate $C_i{}^{(0)}$ into a divergence-free part, $\tilde C_i$ with $\mathcal D\cdot \tilde C=0$, and a pure gradient part according to
\begin{equation}\label{}
C_i{}^{(0)} = \tilde C_i + \partial_i T\,.
\end{equation}
Furthermore, since $C_i{}^{(-1)}=0$, the equations of motion also imply $\partial_u C_i{}^{(0)}=0$, so that $\tilde C_i$ and $T$ can be chosen to be $u$-independent. Equation \eqref{MaxwellSphere} thus reduces to
\begin{equation}\label{}
\left[\Delta - D + 3\right] \tilde C_i=0\,.
\end{equation}
This implies $\tilde C_i=0$ because $\left[\Delta - D + 3\right]$ is invertible and hence that $C_i{}^{(0)}$ is a pure-gauge configuration, $C_i{}^{(0)}=\partial_i T$.

Moving to spin two, we need to discuss 
\begin{align}
& 0 =\left[\Delta-D+2\right] C_{ij}{}^{(-1)}-\frac{D-2}{2(D-3)}\left[\mathcal D_{(i} \mathcal D \cdot C_{j)}{}^{\!(-1)}-\frac{2}{D-2}\, \gamma_{i j}\mathcal D \cdot \mathcal D \cdot C^{(-1)}\right] ,\label{SR-12} \\[5pt]
& (D-4)\, \partial_{u} C_{i j}{}^{(-1)} =[\Delta-2] C_{i j}{}^{(-2)}-\left[\mathcal D_{(i} \mathcal D \cdot C_{j)}{}^{\!(-2)}-\frac{2}{D-2}\, \gamma_{i j}(\mathcal D\cdot)^{2} C^{(-2)}\right] , \label{SR-22}
\end{align}
which are the only two instances of \eqref{dCs} above the radiation order that are not identically satisfied on account of \eqref{zero-Cs} and \eqref{0overoverl}.
Note also that, in view of \eqref{0overoverl},
\begin{equation}
\partial_u C_{ij}{}^{(-2)}=0 \quad \Rightarrow \quad \partial_u^2 C_{ij}{}^{(-1)}=0\,.
\end{equation}
That is, $C_{ij}{}^{(-2)}$ is $u$--independent while $C_{ij}{}^{(-1)}$ is at most linear in $u$:
\begin{equation}\label{Cijnotation}
C_{ij}{}^{(-2)}(\vx) = H_{ij}(\vx)\,,\qquad
C_{ij}{}^{(-1)}(u, \vx) = F_{ij}(\vx) + u\, G_{ij}(\vx)\,.
\end{equation}
We then have
\begin{align}
&0=\left[\Delta-D+2\right] F_{i j}-\frac{D-2}{2(D-3)}\left[\mathcal D_{(i} \mathcal D\cdot F_{j)}-\frac{2}{D-2}\, \gamma_{i j} (\mathcal D\cdot)^2{F}\right]\label{1SR2} ,\\[5pt]
&0=\left[\Delta-D+2\right] G_{i j}-\frac{D-2}{2(D-3)}\left[\mathcal D_{(i} \mathcal D\cdot G_{j)}-\frac{2}{D-2}\, \gamma_{i j}(\mathcal D\cdot)^2{G}\right]\label{2SR2} ,\\[5pt]
&(D-4)\, G_{i j} =\left[\Delta-2\right] H_{i j}-\left[\mathcal D_{(i} \mathcal D \cdot H_{j)}-\frac{2}{D-2}\,\gamma_{i j}(\mathcal D \cdot)^{2}H\right]\label{3SR2} .
\end{align}
The first two relations imply
\begin{equation}\label{GT}
F_{ij}(\vx)=2\left(
\mathcal D_i \mathcal D_j -\frac{1}{D-2}\,\gamma_{ij}\Delta
\right) T(\vx)\,, \quad
G_{ij}(\vx)=\left(
\mathcal D_i \mathcal D_j -\frac{1}{D-2}\,\gamma_{ij}\Delta
\right) S(\vx)\,.
\end{equation}
To see why this is the case, let us focus on the second one and use the decomposition
\begin{equation}\label{Cijdecomp}
G_{ij} = \tilde G_{ij} +  \mathcal D_{(i} \tilde v_{j)} + \left(\mathcal D_i \mathcal D_j -\frac{1}{D-2}\,\gamma_{ij}\Delta\right) S\,,
\end{equation}
where
\begin{equation}\label{}
\mathcal D\cdot \tilde G_i=0\,,\qquad
\gamma^{ij}\tilde G_{ij}=0\,,\qquad
\mathcal D\cdot \tilde v=0\,,
\end{equation}
which is tantamount to the decomposition of the irreducible $so(D-2)$ tensor $G_{ij}$ in irreducible $so(D-3)$ components.
Substituting into the divergence of \eqref{2SR2}, we find
\begin{equation}\label{}
\left[\Delta- D+3\right]\left[\Delta + D-3\right] \tilde v_i=0\,.
\end{equation}
This implies that $\tilde v_i$ belongs to the kernel of $[\Delta+D-3]$, i.e., $\tilde v_i$ is an irreducible harmonic with $\ell=1$ as discussed in Appendix \ref{app:eigenD}. Such vectors give zero contribution to \eqref{Cijdecomp}. Then, from \eqref{2SR2},
\begin{equation}\label{}
\left[\Delta - D + 2 \right] \tilde G_{ij}=0 \quad \Rightarrow \quad \tilde G_{ij}=0 \, .
\end{equation}
This proves \eqref{GT} given \eqref{2SR2}.

The condition in \eqref{3SR2} can be then regarded as an equation for $H_{ij}(\vx)$ given the source term $G_{ij}(\vx)$, or equivalently $S(\vx)$ in view of \eqref{GT}. To solve it, it is convenient to resort to the following decomposition for  the traceless tensor $H_{ij}$, \begin{equation}\label{decompHij}
H_{ij}=\tilde H_{ij} + \mathcal D_{(i} v_{j)} - \frac{2}{D-2}\,\gamma_{ij} \mathcal D\cdot v\,,
\end{equation}
where $\mathcal D \cdot \tilde H_i=0$, while $v_i$ is now a generic vector. 
Substituting \eqref{GT} and \eqref{decompHij} into \eqref{3SR2} and taking divergences eventually allows one to show that
$
\tilde H_{ij}=0
$
and that
\begin{equation}
S = -\frac{2}{D-2}\, \mathcal D\cdot v\,
\end{equation}
up to a constant terms and $\ell=1$ scalar harmonics.
Substituting into \eqref{Cijnotation}, we find that the most general solution for the overleading terms is 
\begin{align}\label{}
C_{ij}^{(-2)}(\vx) &= \mathcal D_{(i} v_{j)}(\vx) - \frac{2}{D-2}\,\gamma_{ij} \mathcal D\cdot v(\vx) \,,\\[5pt]
C_{ij}^{(-1)}(u, \vx) &= 2\left(
\mathcal D_i \mathcal D_j -\frac{1}{D-2}\,\gamma_{ij}\Delta
\right)\left(T(\vx) - \frac{u}{D-2}\, \mathcal D\cdot v(\vx)\right)\,,
\end{align}
where we recognise a supertranslation and a superrotation, parametrised by $T(\vx)$ and $v_i(\vx)$ respectively.

A very similar chain of arguments allows one to prove explicitly that, even for spin-three fields, the only nontrivial overleading components $C_{ijk}^{(-4)}$, $C_{ijk}^{(-3)}$, $C_{ijk}^{(-2)}$ allowed by the equations of motion take precisely the form of the asymptotic symmetries identified in the previous sections.
It is  natural to expect that this pattern actually holds for all spins so that, in this setup, only the structures trivially allowed by the gauge symmetry can appear to overleading orders.

\subsection{Superrotation charges} \label{sec:charges}

We now discuss the structure of surface charges that could be associated to all higher-spin superrotations. To identify it we shall follow, at least in some steps, a strategy similar to that we employed in Section~\ref{hspst} to define finite supertranslation charges. Let us stress, however, that the setup is not completely equivalent, as it is manifest already for spin-two fields. Indeed, evaluating the surface charge \eqref{surface-charge} for all residual symmetries \eqref{symm2} of the Bondi gauge on field configurations with overleading, pure-gauge terms
\begin{align}
	h_{uu} & = - \frac{2 \left( \Delta + D - 2 \right)}{(D-2)^2}\, \Ddot \tilde{v}(\vx) + \cO(r^{1-\frac{D}{2}}) \, , \\
	h_{ui} & = \frac{\cD_i \left( \Delta + D - 2 \right)}{D-2} \left( \tilde{T}(\vx) - \frac{u}{D-2}\, \Ddot \tilde{v}(\vx) \right) + \cO(r^{2-\frac{D}{2}})
\end{align}
one obtains
\begin{align}
	\cQ(u) & = \lim_{r\to\infty} \frac{2\,r^{D-3}}{D-2} \oint d\Omega_{D-2}\! \left\{ T(\vx) \left( \Delta + D-2 \right) \Ddot \tilde{v}(\vx) - \Ddot v(\vx) \left( \Delta + D-2 \right) \tilde{T}(\vx) \right\} \nn \\
	& + \cO(r^{\frac{D-2}{2}}) \, . \label{div_Qrot}
\end{align}
Contrary to the discussion in Section~\ref{hspst}, based on the more restrictive boundary conditions \eqref{falloff_st} only allowing for asymptotic supertranslations, here the pure-gauge overleading field configurations give a divergent contribution to the charge of $\cO(r^{D-3})$. Notice, however, that the surface charge \eqref{div_Qrot} already diverges linearly in four space-time dimensions. In this context, the charge has been regularised in \cite{Lambda-BMS-1,Lambda-BMS-2} (see also \cite{superrotations2, Campiglia:2020qvc}) and in the following we assume that a similar regularisation is possible also for higher values of the space-time dimension $D$. This conjecture will be cross-checked by comparing with the charges that have been used to derive subleading soft theorems from asymptotic symmetries \cite{superrotations,superrotations_anyD}. 

Assuming that the divergence of $\cO(r^{D-3})$ associated to the overleading terms can be cancelled by adding suitable counterterms to the surface charge \eqref{div_Qrot}, one is left with a divergence of $\cO(r^{\frac{D-2}{2}})$ related to radiation. In analogy with Section \ref{hspst}, the latter could be eliminated by defining the charge as the evolution under the equations of motion of $\cQ(-\infty)$ and assuming that no radiation is present for $u$ smaller than a given $u_0$. This would amount to computing the surface charge \eqref{surface-charge} for $u < u_0$ on stationary field configurations, that is on \cite{memory-anyD-Wald}
\begin{subequations}\label{hstationary}
\begin{align}
h_{uu} & = r^{3-D} U^{(0,D-3)}(\vx) + \cO(r^{2-D}) \, , \\[10pt]
h_{ui} & = r^{4-D} U_i{}^{(1,D-4)}(\vx) + r^{3-D} U_i{}^{(1,D-3)} (u,\vx) + \cO(r^{2-D}) \, .
\end{align}
\end{subequations}
Notice that $U_i{}^{(1,D-4)}(\vx)$ is restricted to be $u$-independent and divergence-free on stationary solutions, but it need not vanish.
Substituting these solutions into the charge \eqref{SurfaceST} and including both supertranslations and superrotations for completeness, one gets 
\be \label{charge-stationary}
\begin{split}
\cQ(u) & = - (D-2)  \oint d\O_{D-2}\, T\,U^{(0,D-3)} \\
&+\lim_{r\to\infty} r\, (D-2) \oint d\O_{D-2}\, v^i U_i{}^{(1,D-4)}\\
& - \oint d\O_{D-2}\, v^i \left\{ u\,\pr_i U^{(0,D-3)} - (D-1)\, U_i{}^{(1,D-3)} \right\} .
\end{split} 
\ee 
In the first line we recovered the (finite) supertranslation charge \eqref{SurfaceST} already discussed in Section \ref{hspst}. The second line exhibits instead a linear divergence\footnote{The linear divergence vanishes on shell (as it should) if $v^i$ generates a global symmetry: indeed in this case $U_i{}^{(1,D-4)} \sim \Ddot C_i{}^{(D-5)}$ (see eq.~\eqref{U(k,n)}), while $v^i$ satisfies the conformal Killing equation.} in $r$ involving the superrotation vector $v^i$. In other words, for superrotations, restricting to stationary solutions does not completely solve the issue of the $\cO(r^{\frac{D-2}{2}})$ contributions related to radiation. 

Actually, this singular behaviour in $r$ is not the only puzzling feature of \eqref{charge-stationary}. Even its last line, which is finite in the limit $r\to\infty$ for fixed $u$, seems to diverge as $u$ is then sent to $-\infty$, i.e.\ as one approaches $\mathscr I^+_-$. Indeed the equations of motion for stationary solutions require
\begin{equation}\label{statconstraint}
	u\,\pr_i U^{(0,D-3)}(\vx) - (D-1)\,\left[ U_i{}^{(1,D-3)}(u,\vx)- q_i(\vx)\right] = u \left[\Delta -1\right] U_i^{(1,D-4)}(\vx)\,,
\end{equation}
where $q_i(\vx)$ is an arbitrary $u$-independent integration function, so that  \eqref{charge-stationary} can be recast as
\be \label{charge-stationary-expl}
\begin{split}
	\cQ(u) & = - (D-2)  \oint d\O_{D-2}\, T\,U^{(0,D-3)} + (D-1) \oint d\O_{D-2}\, v^i q_i  \\
	&+\lim_{r\to\infty} r\, (D-2) \oint d\O_{D-2}\, v^i U_i{}^{(1,D-4)}- u \oint d\O_{D-2}\, v^i \left[\Delta -1\right] U_i^{(1,D-4)}\,.
\end{split} 
\ee 
In view of these observations, we note that further restricting to the set of \emph{Coulombic} stationary solutions already considered in \cite{ACD2}, for which the divergence-free component $U_i^{(1,D-4)}$ is zero, namely
\begin{subequations}
	\begin{align}
		h_{uu} & = r^{3-D} U^{(0,D-3)}(\vx) + \cO(r^{2-D}) \, , \\[10pt]
		h_{ui} & = r^{3-D} U_i{}^{(1,D-3)} (u,\vx) + \cO(r^{2-D}) \,,
	\end{align}
\end{subequations}
solves both problems at once. The second line of \eqref{charge-stationary-expl} indeed vanishes identically and we retrieve a well-behaved expression for the charge near $\mathscr I^+_-$,
\begin{equation}\label{charge-stat-finite}
	\cQ\big|_{\mathscr I^+_-} = - (D-2)  \oint d\O_{D-2}\, T(\vx)\,U^{(0,D-3)}(\vx) + (D-1) \oint d\O_{D-2}\, v^i(\vx) q_i(\vx)\,,
\end{equation}
where $q_i(\vx)$ is defined by
\begin{equation}\label{statconstraintC}
  q_i(\vx) = U_i{}^{(1,D-3)}(u,\vx) - \frac{u}{D-1}\, \pr_i U^{(0,D-3)}(\vx) \, .
\end{equation} 
Relabelling $U^{(0,D-3)}\equiv \mathcal M$ and $q_i\equiv \mathcal N_i$, the expression \eqref{charge-stat-finite} for the charge thus 
agrees with the one presented in \cite{ACD2} for global symmetries.

The behaviour of $\mathcal Q(u)$ for generic $u$ can be then retrieved from \eqref{charge-stat-finite} by considering its evolution under the equations of motion, in the same way as we discussed for the supertranslation charge in Section \ref{hspst}. In particular, considering that both $U^{(0,D-3)}$ and $U^{(1,D-3)}$ will acquire a non-trivial $u$-dependence as dictated by the equations of motion in the presence of radiation, rewriting the charge as in \eqref{charge-integral} we find
\be
\begin{split}
\cQ|_{\mathscr I^+_+}-\cQ|_{\scri^+_-} & =
- (D-2) \int_{-\infty}^{+\infty}du  \oint d\O_{D-2}\, T\,\partial_u U^{(0,D-3)}\\
&- \int_{-\infty}^{+\infty} du \oint d\O_{D-2}\, u\, \Ddot v\, \pr_u U^{(0,D-3)} \\
& + \int_{-\infty}^{+\infty} du \oint d\O_{D-2}\, v^i \left\{ \pr_i U^{(0,D-3)} - (D-1) \pr_u U_i{}^{(1,D-3)} \right\} \,.
\end{split}
\ee
If one considers a global, Poincar\'e transformation the right-hand side of this relation vanishes identically, and indeed the surface charge $\mathcal Q(u)$ must be independent of $u$ in this case, even in the presence of radiation \cite{ACD2}. The same result can be achieved by instead restricting the calculation to \emph{static} (i.e.\ $u$-independent), rather than \emph{stationary} or Coulombic solutions (see Appendix~\ref{app:stationary}).
The first line is just the soft supertranslation charge that has been employed in Section \ref{hspst} to derive the leading soft theorem in any even dimension.
The second line exhibits a term linear in $u$ and corresponds to the superrotation charge that has been used in \cite{superrotations_anyD} to derive the subleading soft-graviton theorem in any even dimension \cite{sub-soft_anyD}. Recovering it in our approach supports our conjecture that the regularisation of \cite{Lambda-BMS-1,Lambda-BMS-2} can be extended also to higher space-time dimensions and successfully applied to \eqref{surface-charge}.

We now move to the higher-spin case. All types of divergences encountered above for spin two continue to be present and they are even more severe for $s>2$.\footnote{The overleading pure-gauge contributions to the boundary conditions \eqref{falloffs_superrotations} bring a divergence of $\cO(r^{D+2s-7})$, radiation brings a divergence of $\cO(r^{\frac{D+2s-6}{2}})$, while stationary configurations bring a divergence of $\cO(r^{s-1})$.} Let us observe, however, that for any value of the spin all divergences in $r$ can be eliminated by evaluating the charge on static, rather than stationary, solutions for $u < u_0$ and that this operation gives a consistent surface charge even if we are not dealing with global symmetries. Static solutions are indeed defined by
\be \label{static_def}
\pr_{u} U^{(k,n)} = 0 \quad \forall\ n \, , 
\ee
and we argue in Appendix~\ref{app:stationary} that this condition implies
\be \label{static}
U^{(k,n)} = 0 \quad \textrm{for} \quad n < D-3
\ee 
together with
\be \label{killing}
\cK^{(k)} \equiv \cD\, U^{(k,D-3)} - \frac{2}{D+2(k-2)}\, \g\, \Ddot U^{(k,D-3)} = 0 \, .
\ee
The latter traceless combination is the conformal Killing equation for the rank--$k$ tensor $U^{(k,D-3)}(\vx)$ on the celestial sphere  \cite{eastwood}.

Given \eqref{static}, with this prescription the surface charge is manifestly finite in the limit $r \to \infty$ and we now show that the property \eqref{killing} of the leading contributions of any static solution guarantees that the charge thus defined is also conserved in $u$. Notice that this is not obvious a priori, since the gauge parameters generating superrotations bring a polynomial dependence in $u$ (cf.~\eqref{ansatz-parameter} or, e.g., \eqref{symm2} and \eqref{symm3}):
\be \label{sr-charge_1}
\cQ_{\textrm{static}}(u) = \oint d\Omega_{D-2} \sum_{k=0}^{s-1} (-1)^{s+k} \binom{s-1}{k} (D+s+k-4)\, \rho^{(k)}(u,\vx) U^{(k,D-3)}(\vx) \, .
\ee
A dependence on $u$ could imply that the charge is not conserved even in the absence of radiation and would also create problems in evaluating it for $u \to \pm \infty$. The $u$-dependence of the rank--$k$ traceless tensors $\rho^{(k)}$ is however fixed by \eqref{diff_u} as 
\be
\rho^{(k)}(u,\vx) = K^{(k)}(\vx) + \sum_{m=1}^{s-k-1} \frac{(-1)^m\,u^m}{(D+s+k-4)_m} \binom{s-k-1}{m} (\Ddot)^m K^{(k+m)}(\vx) \, ,
\ee
where, for a field of spin $s$, we introduced the $s-1$ traceless tensors on the celestial sphere $K^{(0)}$, $K^{(1)}_i$,\ldots, $K^{(s-1)}_{i_{s-1}}$ (for $s = 3$ they correspond to $K^{(0)} = T$, $K^{(1)}_i=\rho_i$ and $K^{(2)}_{ij} = K_{ij}$ in the notation of \eqref{parameters3}).
When the field configuration is static, each term in the charge \eqref{sr-charge_1} which depends on $u$ thus contains at least one divergence of the tensors $K^{(l)}$. Integrating it by parts one reconstructs the conformal Killing equation for the tensors $U^{(k,D-3)}$ and thus the contribution vanishes on shell on account of \eqref{killing}. The surface charge associated to higher-spin superrotations thus becomes 
\be \label{sr-charge}
\cQ_{\textrm{static}} = \oint d\Omega_{D-2} \sum_{k=0}^{s-1} (-1)^{s+k} \binom{s-1}{k} (D+s+k-4)\, K^{(k)}(\vx) U^{(k,D-3)}(\vx) \, .
\ee
The result has the same form as the charge associated to global higher-spin symmetries \cite{ACD2}, with the difference that now the traceless tensors $K^{(k)}$ do not satisfy any differential equation (e.g., for $s=3$ they need not obey \eqref{diff_constr}). In analogy with the discussion in Section~\ref{hspst}, a contribution from radiation might be obtained by substituting in \eqref{sr-charge} the  $U^{(k,D-3)}(u,\vx)$ obtained via the evolution under the equations of motion of a field configuration that is static for $u$ smaller than a given $u_0$. 
It would be obviously preferable, however, to implement a procedure leading to finite charges for more general configurations. 
We hope to come back to this issue in future work, with the goal of finding a proper renormalisation scheme valid in higher dimensions and for any value of the spin.

\acknowledgments

We are grateful to G.~Barnich, D.~Grumiller and M.~M.~Sheikh-Jabbari for discussions.
The work of A.C.\ was supported by the Fonds de la Recherche Scientifique - FNRS under Grants No.\ F.4503.20 (``HighSpinSymm'') and T.0022.19 (``Fundamental issues in extended gravitational theories'').
D.F. gratefully acknowledges the Asian-Pacific Center for Theoretical Physics (APCTP) in Pohang for the kind hospitality extended to him during the preparation of this work.
The work of C.H. is supported by the Knut and Alice Wallenberg Foundation under grant KAW 2018.0116.

\appendix

\section{Notation and conventions} \label{app:notation}

Throughout the paper we employ retarded Bondi coordinates  $(x^\mu)=(u,r,x^i)$, where $x^i$, for $i=1,2,\ldots,n$, denotes~the $n := D-2$ angular coordinates on the sphere at null infinity. In these coordinates, the Minkowski metric reads 
\be \label{bondi-coord}
ds^2 = - du^2 - 2 du dr + r^2 \gamma_{ij}\, dx^i dx^j\,,
\ee
where $\gamma_{ij}$ is the metric of the Euclidean $n$-sphere $S^n$. We denote by $\cD$ the covariant derivative on the sphere while $\cD \cdot$ and  $\Delta$ stand for the corresponding divergence and Laplacian, respectively.

We use a symbol with a subscript, like $\m_k$ or $i_k$, to denote a group of symmetrised indices, whose number is specified by the subscript $k$. Repeated indices denote instead a symmetrisation without any overall factor. For instance, $\nabla_\m \e_{\m_{s-1}}$ is a shorthand for $\nabla_{\!\m_1} \e_{\m_2 \cdots \m_{s}} + \nabla_{\m_2} \e_{\m_3 \cdots \m_s \m_1} + \cdots$. In Section~\ref{hspsr} and in Appendices~\ref{app:symm} and \ref{app:stationary}, we also employ an index-free notation, where all symmetrised indices are omitted and the trace is denoted by a prime. In this case, the previous expression is denoted by $\nabla \e$.

\section{Geometry of the sphere and polarisations}
\label{app:sphereandpol}

\subsection{Properties of the $n$-sphere}
\label{app:geometry}
Let us recall here some properties of the embedding of the unit $n$-sphere $S^n$ in the Euclidean space $\mathbb R^{n+1}$. Changing coordinates according to $x^I=r\, \hat x^I$ where $x^I$ are Cartesian coordinates on $\mathbb R^{n+1}$ and $\hat x^I$ is a parametrisation of unit vectors in terms of the angles $x^i$, the Euclidean metric reads 
\begin{equation}
ds^2 = dx^I\,dx^I = dr^2 + r^2 \gamma_{ij} dx^i dx^j\,,
\end{equation}
where 
\begin{equation}\label{metricintsphere}
\gamma_{ij}=e_i^I\, e_j^I\,,\qquad e_i^I=\partial_i \hat x^I\,.
\end{equation}

The induced metric on the unit sphere $\gamma_{ij}$ also defines a covariant derivative $\mathcal D_i$ thereon.
As a consequence of $\partial_I \partial_J x^K=0$, one can show the useful identity
\begin{equation}
\label{basicspherid1}
\mathcal D_i  \mathcal D_j {\hat x}^I + \gamma_{ij} {\hat x}^I=0\,.
\end{equation}
The metric $\gamma_{ij}$ can be also represented by the rank-$n$ matrix 
\begin{equation}\label{metricextsphere}
\gamma_{IJ} = \delta_{IJ}-{\hat x}_I  {\hat x}_J
\end{equation}
which projects any vector to its component tangent to the sphere. Therefore, for any unit vector ${\hat q}^I$,  its projection on the sphere ${\hat q}_i={\hat q}^I e_i^I$ obeys 
\begin{equation}\label{basicspherid2}
 \hat q_i \gamma^{ij} \hat q_j= 1-({\hat x}^I {\hat q}^I)^2\,.
\end{equation}

\subsection{Spectrum of $\Delta$}
\label{app:eigenD}

The eigenvalues $\lambda$ of the operator $\Delta = \mathcal D_i \mathcal D^i$ acting on symmetric, traceless, divergence-free tensors $K_{i_1i_2\cdots i_s}$, i.e. irreducible tensors of rank $s$,
\begin{equation}\label{eigenfunctions}
  (\Delta - \lambda)K_{i_1i_2\cdots i_s}=0\,,
\end{equation} 
are
\begin{equation}\label{eigenvalues}
\lambda=-\ell(\ell+n-1)+s\,,\qquad \ell= s, s+1,s+2,\ldots\ .
\end{equation}
Explicit eigenfunctions are provided by the irreducible tensor spherical harmonics. They can be constructed starting from tensors with constant Cartesian components 
\begin{equation}
C_{[A_1I_1][A_2I_2]\cdots [A_sI_s]I_{s+1}\cdots I_\ell}\qquad \text{for }\ell\ge s\,,
\end{equation} 
which are assumed to be
completely traceless,
symmetric under permutations of $I_{s+1},\ldots,I_\ell$,
symmetric under permutations of the pairs $[A_kI_k]$, and antisymmetric under exchanges of an $A_k$ with its corresponding $I_k$ index.
It then follows that the tensors
\begin{equation}\label{}
K_{A_1A_2\cdots A_s} = C_{[A_1I_1][A_2I_2]\cdots [A_sI_s]I_{s+1}\cdots I_\ell}\ x^{I_1} x^{I_2}\cdots x^{I_\ell}
\end{equation}
are symmetric, harmonic $\Delta_{\mathbb R^{n+1}}K_{A_1 A_2\cdots A_s}=0$ and homogeneous of degree $\ell$ under rescalings of $x^I$. They are also tangent to the sphere $x^A K_{A A_2\cdots A_s}=0$ and the corresponding tensors $K^{i_1i_2\cdots i_s}$ on the sphere, defined by
\begin{equation}\label{}
K^{A_1 A_2\cdots A_s}=r^{\ell} e_{i_1}^{A_1} e_{i_2}^{A_2}\cdots e_{i_s}^{A_s} K^{i_1i_2\cdots i_s}\,,
\end{equation}
are divergence-free and trace-free, as can be checked using \eqref{metricintsphere}, \eqref{basicspherid1} and \eqref{metricextsphere}, together with the above properties.
Using also 
\begin{equation}\label{}
\Delta_{\mathbb R^{n+1}}f
= \frac{1}{r^n}\partial_r\left(r^n\partial_r f\right) + \frac{1}{r^2}\,\Delta f\,,
\end{equation}
together with the properties of $K^{i_1i_2\cdots i_s}$ and the action \eqref{basicspherid1}, one then obtains
\begin{equation}\label{}
0=\Delta_{\mathbb R^{n+1}} K^{A_1 A_2\cdots A_s} = r^{\ell-2} e_{i_1}^{A_1} e_{i_2}^{A_2}\cdots e_{i_s}^{A_s} \left[\Delta+\ell(\ell+n-1)-s\right]K^{i_1i_2\cdots i_s}\,,
\end{equation}
thus retrieving \eqref{eigenfunctions} with the eigenvalues \eqref{eigenvalues}.
The uniqueness of these eigenvalues can be inferred from the density of the homogeneous polynomials in the domain of $\Delta$ under consideration. 

The spectrum on reducible tensors can be obtained from the above one by first decomposing the desired tensor in terms of symmetrised gradients of irreducible tensors
and then using the commutation relation
\begin{equation}\label{commRiemann}
[\mathcal D_i, \mathcal D_j]v^k = R\indices{^k_{lij}} v^l\,,\qquad
R_{ijkl}=\gamma_{ik}\gamma_{jl}-\gamma_{il}\gamma_{jk}\,.
\end{equation}
For instance, for a traceless but generically not divergence free tensor $C^{(s)}_{i_1i_2\cdots i_s}$, the desired decomposition can be cast in the form
\begin{equation}\label{}
C^{(s)}_{i_1i_2\cdots i_s} = K^{(s)}_{i_1i_2\cdots i_s} + \mathcal D_{(i_1}K^{(s-1)}_{i_2\cdots i_s)} + \mathcal D_{(i_1} \mathcal D_{i_2} K^{(s-2)}_{i_3\cdots i_s)}+\cdots + \mathcal D_{(i_1}\mathcal D_{i_2}\cdots \mathcal D_{i_s)}K^{(0)}+\cdots\,,
\end{equation} 
where the dots implement a traceless projection.
More details for $s=0,1,2$ are available in \cite{Rubin:1983be} for  irreducible case and in \cite{Rubin:1984tc} for the reducible case.

\subsection{Polarisation Tensors}
\label{app:polarizations}

In this appendix we sketch the construction of a useful set of polarization tensors. We apply the notation of Appendix \ref{app:geometry} for quantities defined on the $n=D-2$ sphere.

To construct the physical polarisations for the electromagnetic potential $\mathcal A_\mu$, we start from the Fierz system written in Minkowski coordinates,
\begin{equation}
\Box \mathcal A_\mu=0\,,\qquad \partial\cdot \mathcal A=0\,,\qquad \Box \Lambda=0\,,
\end{equation}
where $\Lambda$ is the gauge parameter. According to the first equation, the Fourier transform $\epsilon_\mu$ of $\mathcal A_\mu$ has support restricted to null vectors 
\begin{equation}\label{nullvector}
q^\mu 
=
\omega(1,{\hat x}^I)
\end{equation}
and satisfies, letting $\epsilon_\mu = (\epsilon_0, \epsilon_I)$,
\begin{equation}\label{firstconstr}
\epsilon_0 = -{\hat x}^I  \epsilon^I\,.
\end{equation}
The residual gauge parameters must also have support on vectors of the form \eqref{nullvector} and thus $\epsilon_\mu$ is equivalent, up to a gauge transformation, to
\begin{equation}
\epsilon_\mu \,\sim\, \epsilon_\mu-i\omega(-1,{\hat x}^I)\lambda\,,
\end{equation}
where $\lambda$ stands for the Fourier transform of $\Lambda$.
Taking \eqref{firstconstr} into account and choosing a gauge parameter such that $i\omega \lambda = {\hat x}^I \epsilon^I$ thus leads to 
\begin{equation}
\epsilon_\mu \,\sim\, (0, \gamma_{IJ}\epsilon^J)
\end{equation}
with $\gamma_{IJ}$ as in \eqref{metricextsphere}.
Therefore, $\epsilon_\mu$ can be parametrised by the projection on the unit sphere of a generic vector $\epsilon^I$ and has $D-2=n$ independent components, up to gauge equivalence.

Given a set of coordinates $x^i$ on the sphere, we can choose the following basis for physical polarisations,
\begin{equation}
\epsilon^{(i)}_\mu({\hat x}) = (0,D^i{\hat x}^I).
\end{equation}
In retarded components, one then finds
\begin{equation}\label{polspin1}
\epsilon^{(i)}_u=0=\epsilon^{(i)}_r\,,\qquad
\epsilon^{(i)}_j=\delta^i_j\,.
\end{equation}
With this choice all sums over polarisations can be understood as contractions with the metric $\gamma_{ij}$,
while the projection over the space of physical polarisations is just the projection on the sphere.

The spin-two Fierz system,
\setlength\arraycolsep{2pt}
\begin{equation}
\begin{array}{rl}
\Box h_{\mu\nu} &= 0\,,\\
\Box \Lambda_{\mu} &= 0\,,
\end{array}
\qquad
\begin{array}{rl}
\partial\cdot h_\mu &=0\,,\\
\partial\cdot \Lambda &=0
\end{array}
\qquad
\begin{array}{rl}
h' &=0\,,\\
&
\end{array}
\end{equation}
implies that the Fourier transforms $\epsilon_{\mu\nu}$ and $\lambda_\mu$ of $h_{\mu\nu}$ and $\Lambda_\mu$ have support on the null vectors \eqref{nullvector}, and $\lambda_\mu=(\lambda_0, \lambda_I)$ satisfies
\begin{equation}
\lambda_0 = -{\hat x}^I \lambda^I\,.
\end{equation}
Furthermore $\epsilon_{\mu\nu}$ satisfies
\begin{equation}
\epsilon_{0I}=-{\hat x}^I \epsilon_{IJ}\,,\qquad
\epsilon_{00}={\hat x}^I  \epsilon_{IJ} {\hat x}^J=\delta^{IJ}\epsilon_{IJ}\,.
\end{equation}
The latter condition implies that $\epsilon_{IJ}$ has a traceless projection on the sphere, $\gamma^{IJ}\epsilon_{IJ}=0$.
Up to gauge equivalence, the polarization tensor reads
\begin{align}
\label{gauge00}
\epsilon_{00}
&\,\sim\,
{\hat x}^I  \epsilon_{IJ} {\hat x}^J -2i\omega {\hat x}^I \lambda^I \,, \\
\label{gauge0I}
\epsilon_{0I}
&\,\sim\,
- \epsilon_{IJ}{\hat x}^J + i\omega(\lambda_I +  {\hat x}_I {\hat x}_J  \lambda^J) \,, \\
\label{gaugeIJ}
\epsilon_{IJ}
&\,\sim\,
\epsilon_{IJ}-i\omega({\hat x}_I \lambda_J + {\hat x}_J \lambda_I)\,.
\end{align}
We parametrise the gauge vector as follows,
\begin{equation}\label{parvec}
\lambda^I
=
A\, {\hat x} + B^i e_i^I\,,
\end{equation}
with $e_i^I$ as in \eqref{metricintsphere}.
Imposing that the right-hand sides of \eqref{gauge00}
and \eqref{gauge0I} vanish, we have
\begin{equation}\begin{split}\label{toimpose2}
{\hat x}^I  \epsilon_{IJ} {\hat x}^J-2i\omega A &=0\,,\\
-\epsilon_{IJ} {\hat x}^J + 2i\omega {\hat x}_I A + i\omega  e^i_{I} B_i &=0\,,
\end{split}\end{equation} 
where the index $i$ is raised and lowered using $\gamma_{ij}$\,.
This fixes the coefficients $A$ and $B^i$ to be
\begin{equation}\label{resvec}
2i\omega A = {\hat x}^I  \epsilon_{IJ} {\hat x}^J\,,
\qquad
i \omega B_i =  e_i^I \epsilon_{IJ} {\hat x}^J\,,
\end{equation}
upon taking suitable projections of the second equation in \eqref{toimpose2}.
One thus finally arrives at the expression 
\begin{equation}
\epsilon_{00}\,\sim\, 0\,,\qquad
\epsilon_{0I}\,\sim\, 0\,,\qquad
\epsilon_{IJ}\,\sim\,\gamma_{IK}\,\gamma_{JL}\,\epsilon_{KL}
\end{equation}
substituting \eqref{parvec} and \eqref{resvec} into \eqref{gaugeIJ}.
Recalling the constraint $\gamma^{IJ}\epsilon_{IJ}=0$, this implies that the most general polarization tensor, up to gauge equivalence, is identified by a symmetric traceless tensor on the $(D-2)$-sphere and thus characterises $\frac{(D-2)(D-1)}{2}-1$ degrees of freedom.

A convenient basis for symmetric tensors on the sphere is furnished by
\begin{equation}
E_{kl}^{(ij)}
= \delta^{i}_k \delta^{j}_l + \delta^{i}_l \delta^{j}_k\,,
\end{equation}
whose traceless projection reads
\begin{equation}\label{polspin2}
2 \epsilon^{(ij)}_{kl} = \delta^{i}_k \delta^{j}_l + \delta^{i}_l \delta^{j}_k - \frac{2}{{D-2}} \gamma^{ij}\gamma_{kl}\,.
\end{equation}
We therefore adopt polarization tensors $\epsilon^{(ij)}_{\mu\nu}$ such that, in retarded components, $\epsilon^{(ij)}_{uu}=0$, $\epsilon^{(ij)}_{ur}=0$, $\epsilon^{(ij)}_{uk}=0$, $\epsilon^{(ij)}_{rr}=0$, $\epsilon^{(ij)}_{rk}=0$ and with angular components $\epsilon^{(ij)}_{kl}$ specified by equation \eqref{polspin2}.

We consider the spin-three Fierz system
\setlength\arraycolsep{2pt}
\begin{equation}
\begin{array}{rl}
\Box \varphi_{\mu\nu\rho} &= 0\,,\\
\Box \Lambda_{\mu\nu} &= 0\,,
\end{array}
\qquad
\begin{array}{rl}
\partial\cdot \varphi_{\mu\nu} &=0\,,\\
\partial\cdot \Lambda_\mu &=0
\end{array}
\qquad
\begin{array}{rl}
\varphi'_\mu &=0\,,\\
\Lambda' &=0\,.
\end{array}
\end{equation}
The Fourier transforms $\epsilon_{\mu\nu\rho}$ and $\lambda_{\mu\nu}$ of $\varphi_{\mu\nu\rho}$ and $\Lambda_{\mu\nu}$ have support on the null vectors \eqref{nullvector}. The gauge parameter satisfies
\begin{equation}
\lambda_{0I}=-{\hat x}^I \lambda_{IJ}\,,\qquad
\lambda_{00}={\hat x}^I  \lambda_{IJ} {\hat x}^J=\delta^{IJ}\lambda_{IJ}\,,
\end{equation}
so that in particular $\gamma^{IJ}\lambda_{IJ}=0$.
The polarization is instead constrained by 
\begin{equation}\begin{split}
\epsilon_{0IJ} &=- \epsilon_{IJK}\, {\hat x}^K\,,\\
\epsilon_{00I} &= \epsilon_{IJK}\, {\hat x}^I{\hat x}^K\,,\\
\epsilon_{000} &=- \epsilon_{IJK}\, {\hat x}^I{\hat x}^J{\hat x}^K\,,
\end{split}\end{equation}
and satisfies the trace conditions
\begin{equation}
\gamma^{IJ}\epsilon_{0IJ}=0\,,\qquad \gamma^{IJ}\epsilon_{IJK}=0\,.
\end{equation}
Up to gauge equivalence, the polarization tensor reads
\begin{align}
\label{gauge000}
\epsilon_{000}
&\,\sim\,
- \epsilon_{IJK} {\hat x}^I{\hat x}^J{\hat x}^K + 3 i\omega {\hat x}^I {\hat x}^J \lambda_{IJ} \,, \\
\label{gauge00K}
\epsilon_{00K}
&\,\sim\,
{\hat x}^I {\hat x}^J \epsilon_{IJK} - 2i\omega {\hat x}^I \lambda_{IK} - i\omega {\hat x}_K\, \left( {\hat x}^I \epsilon_{IJ}{\hat x}^J \right) \,, \\
\label{gauge0IJ}
\epsilon_{0IJ}
&\,\sim\,
-\epsilon_{IJK}{\hat x}^K+ i\omega \lambda_{IJ} + i\omega({\hat x}_I \lambda_{JK}+{\hat x}_J \lambda_{IK}){\hat x}^K\,,
\\
\label{gaugeIJK}
\epsilon_{IJK}
&\,\sim\,
\epsilon_{IJK}-i\omega({\hat x}_I \lambda_{JK}+{\hat x}_J \lambda_{KI}+{\hat x}_K \lambda_{IJ})\,.
\end{align}
We parametrise the gauge tensor as
\begin{equation}\label{partens}
\lambda_{IJ}
=
A {\hat x}_I {\hat x}_J + B_i( {\hat x}_I  e^i_J + {\hat x}_J  e^i_I) + C_{ij} e^i_I  e^j_J\,,
\end{equation}
where $C_{ij}$ are symmetric traceless coefficients. Imposing that the right-hand sides of \eqref{gauge000},
\eqref{gauge00K} and \eqref{gauge0IJ} vanish then fixes the coefficients $A$, $B_i$ and $C_{ij}$ to be
\begin{equation}\label{restens}
3i\omega A = \epsilon_{IJK}\, {\hat x}^I {\hat x}^J {\hat x}^K\,,
\qquad
2i \omega B_i =  e_i^I \epsilon_{IJK}\, {\hat x}^J {\hat x}^K\,,
\qquad
i\omega C_{ij} =  e_i^I  e_j^J \epsilon_{IJK}\, {\hat x}^K\,.
\end{equation}
One thus arrives at 
\begin{equation}
\epsilon_{000}\,\sim\, 0\,,\qquad
\epsilon_{00I}\,\sim\, 0\,,\qquad
\epsilon_{0IJ}\,\sim\,0\,,\qquad
\epsilon_{IJK}\,\sim\,\gamma_{IL}\,\gamma_{JM}\,\gamma_{KN}\,\epsilon_{LMN}\,.
\end{equation}
substituting into \eqref{gaugeIJK}.
Recalling the constraint $\gamma^{IJ}\epsilon_{IJK}=0$, this means that the most general polarization tensor, up to gauge equivalence, is identified by a symmetric traceless tensor on the $n$-sphere and thus characterises $\frac{({D-2})(D-1)D}{3!}-(D-2)$ degrees of freedom.

A convenient basis for such tensors is furnished by
\begin{equation}\label{polspin3}
3!\, \epsilon^{(ijk)}_{lmn} = \delta^{i}_{(l} \delta^{j}_{m} \delta^{k}_{n)}  - \frac{2}{D}\gamma_{(lm}\gamma^{(ij}\delta^{k)}_{n)}\,.
\end{equation}
We therefore adopt polarization tensors whose only nonzero components are the one with indices on the sphere $\epsilon^{(ij)}_{kl}$ specified by equation \eqref{polspin3}.

For spin-$s$, the Fierz system reads
\setlength\arraycolsep{2pt}
\begin{equation}
\begin{array}{rl}
\Box \varphi_{\mu_{s}} &= 0\,,\\
\Box \Lambda_{\mu_{s-1}} &= 0\,,
\end{array}
\qquad
\begin{array}{rl}
\partial\cdot \varphi_{\mu_{s-1}} &=0\,,\\
\partial\cdot \Lambda_{\mu_{s-2}} &=0
\end{array}
\qquad
\begin{array}{rl}
\varphi'_{\mu_{s-2}} &=0\,,\\
\Lambda'_{\mu_{s-3}} &=0\,.
\end{array}
\end{equation}
As in the previous cases, the transversality conditions in Fourier space imply that one can always trade a $0$ index for a projection along ${\hat x}$, namely
\begin{equation}
\epsilon_{0_m I_{s-m}}=- \epsilon_{0_{m-1}I_{s-m}J}\,{\hat x}^J\,.
\end{equation}
This fact allows us to take the independent components of the gauge parameter and of the polarization tensor to be contained in $\lambda_{I_{s-1}}$ and $\epsilon_{I_s}$ and the trace conditions imply that such tensors have traceless projections on the sphere. Parametrising $\lambda_{I_{s-1}}$ as
\begin{equation}
\lambda_{IJ\cdots K} 
= 
A {\hat x}_{I} {\hat x}_J \cdots {\hat x}_K
+
B_i  e^i_{(I} {\hat x}_J \cdots {\hat x}_{K)}
+
\cdots
+
C_{ij\cdots k}  e^i_{(I} e^j_{J}\cdots  e^k_{K)}\,,
\end{equation} 
with suitable traceless coefficients, one can then solve for
\begin{equation}
\epsilon_{0_mI_{s-m}}\,\sim\,0\,,
\qquad m\ge1\,,
\end{equation}
which maps $\epsilon_{I_s}$ to its projection on the sphere,
\begin{equation}
\epsilon_{IJ\cdots K}\ =\ \gamma_{IL}\gamma_{JM}\cdots \gamma_{KN} \, \epsilon_{LM\cdots N}\,.
\end{equation}
Symmetric traceless tensors on the $n$-sphere indeed possess $\binom{D-3+s}{s}-\binom{D-5+s}{s-2}$ degrees of freedom.

\section{Symmetries of the Bondi-like gauge} \label{app:symm}

In this appendix we provide an algorithm to fix the structure of the subleading terms in the radial expansion of the components $\e_{r_{s-k-1}i_k}$ of the gauge parameter, while checking the consistency of the solution with all constraints coming from the Bondi-like gauge.

In Bondi coordinates, the gauge variation of a generic field component reads
\be \label{variation-s}
\begin{split}
\d \vf_{r_{s-k-l}u_l} & = l\, \pr_u \e_{r_{s-k-l}u_{l-1}} + \frac{s-k-l}{r} \left( r\pr_r - 2k \right) \e_{r_{s-k-l-1}u_l} \\
& + \cD \e_{r_{s-k-l}u_l} + 2\,r\, \g \left( \e_{r_{s-k-l+1}u_{l}} - \e_{r_{s-k-l}u_{l+1}} \right) ,
\end{split}
\ee
where, as in Section~\ref{sec:symm_bondi}, we omitted all sets of symmetrised angular indices.

Focussing on the variation of the components without any index $u$ and using the trace constraint on the gauge parameter, implying
\be
\e_{r_{s-k}u} = \frac{1}{2} \left( \e_{r_{s-k+1}} + \frac{1}{r^2}\, \e'{}_{\!\!r_{s-k-1}} \right) ,
\ee
one finds \eqref{variation-radial-s}:
\be \label{variation-radial-s_app}
\d \vf_{r_{s-k}} = \frac{1}{r} \left\{ (s-k) \left( r\pr_r - 2k \right) \e^{(k)} - \g\, \e^{(k)\prime} \right\} + \cD \e^{(k-1)} + r\,\g\, \e^{(k-2)} = 0 \, .
\ee
Let us recall that we denoted by $\e^{(k)}$ the components $\e_{r_{s-k-1}i_k}$ of the gauge parameter, while a prime denotes a contraction with $\g^{ij}$.
Looking at \eqref{variation-radial-s_app}, it is clear that the general solution of this equation has the form \eqref{ansatz-parameter}, 
\be \label{ansatz-parameter_app}
\e^{(k)}(r,u,\vx) = r^{2k} \rho^{(k)}(u,\vx) + \sum_{l\,=\,k}^{2k-1} r^l \a^{(k,l)}(u,\vx) \, ,
\ee
where the $\a^{(k,l)}$ are determined recursively (and algebraically) in terms of the $\rho^{(l)}$ with $l < k$. Indeed, $\e^{(k-1)}$ and $\e^{(k-2)}$ are fixed by the previous iterations, while the traces of each $\a^{(k,l)}$ can be eliminated by computing the traces of \eqref{variation-radial-s_app}:
\begin{align}
& \frac{1}{r} \left\{ \left[ (k-s) \left( r\pr_r - 2k \right) + m(D+2(k-m-2)) \right] \e^{(k)[m]} + \g\, \e^{(k)[m+1]} \right\} \label{trace_var} \\
& = 2m\, \Ddot \e^{(k-1)[m-1]} + \cD \e^{(k-1)[m]} + r \left[ m(D+2(k-m-2))\, \e^{(k-2)[m-1]} + \g\, \e^{(k-2)[m]} \right] , \nn 
\end{align}
where $\e^{(k)[l+1]} \equiv \g\cdot \e^{(k)[l]}$. For instance, for $k$ even, introducing for simplicity $\a^{(k,2k)} \equiv \rho^{(k)}$, the last available trace gives
\be
\begin{split}
&\left[ (2k-l)(s-k)+\frac{k(D+k-4)}{2} \right] \a^{(k,l)[\frac{k}{2}]} \\
& \hspace{15pt} = k\, \Ddot \a^{(k-1,l-1)[\frac{k-2}{2}]} + \frac{k(D+k-4)}{2}\, \a^{(k-2,l-2)[\frac{k-2}{2}]} \, ,
\end{split}
\ee
and one can substitute the result in \eqref{trace_var} to determine $\a^{(k,l)[\frac{k-2}{2}]}$ and proceed recursively.

The recursion relations become rather cumbersome after the first few values of $k$, but, as an example, we can provide their explicit solution up to $k=3$ (with $s$ generic), that suffices to capture all residual symmetries of the Bondi-like gauge for fields of spin $s \leq 4$:
\begin{align}
\e_{r_{s-1}} & = \rho^{(0)} \, , \\[10pt]
\e_{r_{s-2}} & = r^2 \rho^{(1)} + \frac{r}{s-1}\, \pr \rho^{(0)} \, , \\[5pt]
\e_{r_{s-3}} & = r^4 \rho^{(2)} + \frac{r^3}{s-2} \left( \cD \rho^{(1)} - \frac{2}{D+s-4}\, \g\, \Ddot \rho^{(1)} \right) \nn \\
& \hspace{-15pt} + \frac{(s-3)!\, r^2}{2(s-1)!} \left( \cD^2 \rho^{(0)} - \frac{2}{D+2s-6}\, \g \left[ \Delta - (s-1)(s-2) \right] \rho^{(0)} \right) , \\[5pt]
\e_{r_{s-4}} & = r^6 \rho^{(3)} + \frac{r^5}{s-3} \left( \cD \rho^{(2)} - \frac{2}{D+s-3}\, \g\, \Ddot \rho^{(2)} \right) + \frac{(s-4)!\, r^4}{2(s-2)!}\, \bigg( \cD^2 \rho^{(1)} \nn \\
& \hspace{-15pt} - \frac{2}{D+2s-6}\, \g \Big[ \left( \Delta + D - s(s-5) - 9 \right) \rho^{(1)} + \frac{2(D+2s-7)}{D+s-4}\, \cD \Ddot \rho^{(1)} \Big] \bigg) \nn \\
& \hspace{-15pt} + \frac{(s-4)!\,r^3}{6(s-1)!} \left( \cD^3 \rho^{(0)} - \frac{2}{D+2s-6}\, \g\, \cD \left( 3\Delta + 2D - s(3s-13) - 18 \right) \rho^{(0)} \right) . 
\end{align}
Setting to zero all $\rho^{(k)}$ with $k > 0$, one recovers the full structure of the supertranslation parameters (cf.\ \eqref{synthST}) up to $k=3$ for any value of $s$. 

The $\e^{(k)}$ are the only components of the gauge parameter that enter the surface charge \eqref{surface-charge}. Still, we imposed only part of the conditions necessary to preserve the Bondi-like gauge \eqref{bondig} and we have to check if the rest imposes additional constraints on the tensors $\rho^{(k)}$. Preserving the vanishing of the traces of the non-zero components of the field yields 
\be \label{var_trace_u}
\begin{split}
\d \vf'{}_{\!u_{s-k}} & = -\,2r \left[ (D+2(k-3)) \e_{u_{s-k-1}} + \g\, \e'{}_{\!u_{s-k+1}} \right] + (s-k)\, \pr_u \e'{}_{\!u_{s-k-1}} \\[5pt]
& + 2\, \Ddot \e_{u_{s-k}} + \cD \e'{}_{\!u_{s-k}} + 2r \left[ (D+2(k-3)) \e_{ru_{s-k}} + \g\, \e'{}_{\!ru_{s-k}} \right] = 0 \, ,
\end{split}
\ee
which fixes the components $\e_{u_{s-k-1}}$ using the same strategy we employed in the analysis of \eqref{variation-radial-s}. Indeed, all components with at least one index $r$ and one index $u$, like those appearing in the last line of \eqref{var_trace_u}, can be rewritten in terms of the $\e^{(k)}$ using the trace constraint on the gauge parameter:
\be
\x_{r_{\a}u_{\b}} = \frac{1}{2} \left( \x_{r_{\a+1}u_{\b-1}} + \frac{1}{r^2}\, \x'{}_{r_{\a-1}u_{\b-1}} \right) .
\ee

At this stage we have fixed all components of the gauge parameter in terms of the tensors $\rho^{(0)}(u,\vx)$, \ldots, $\rho^{(s-1)}(u,\vx)$, but we still have to check if the vanishing of the variations \eqref{variation-s} with $l \geq 1$ and $k+l<s$ imposes additional constraints on them. This question has been already addressed in Section~5.2 of \cite{ACD2}, where it has been shown that preserving the Bondi-like gauge requires \eqref{diff_u}, that is
\be
\pr_u \rho^{(k)} + \frac{s-k-1}{D+s+k-4}\, \Ddot \rho^{(k+1)} = 0 \quad \textrm{for} \quad k < s-1 \, .
\ee
%

\section{Stationary and static solutions of Fronsdal's equations} \label{app:stationary}

In this appendix we characterise the behaviour near $\scri^+$ of stationary and static solutions of Fronsdal's equations, assuming that asymptotically the fields can be expanded in powers of the radial coordinate as in \eqref{exp-spin-s}. Under this hypothesis, the source-free Fronsdal equations $\cF_{r_{s-k}i_k} = 0$ imply
\be \label{exp-Frs}
(n+2k)(D-n-3)\, U^{(k,n)} = (n+2k-1)\, \Ddot U^{(k+1,n-1)} \, ,
\ee
while the equations $\cF_{u_{s-k}i_k} = 0$ give
\be \label{exp-Fus}
\begin{split}
& \left[ (D-n-2)(s-k-1) + (n+2k) \right] \pr_u U^{(k,n)} = (s-k)\, \pr_u \Ddot U^{(k+1,n-1)} \\[5pt]
& \hspace{15pt} - \left[ \Delta - (n-1)(D-n-2k-2) - k(D-k-2) \right] U^{(k,n-1)} + \cD \Ddot U^{(k,n-1)} \\[5pt]
& \hspace{15pt} - (D-n-4)\, \cD U^{(k-1,n)} - 2\,\g\,\Ddot U^{(k-1,n)} + 2(D-n-4)\,\g\,U^{(k-2,n+1)} \, .
\end{split}
\ee
All other equations of motion are identically satisfied in the Bondi-like gauge. Combining \eqref{exp-Frs} with the traces of \eqref{exp-Fus} one obtains \eqref{U(k,n)}, and using the latter in \eqref{exp-Fus} with $k=s$ one eventually obtains \eqref{dCs}. In the following, instead, we shall also have to consider the equations above for values of $n$ for which some of these substitutions are not possible. For instance, for $n=D-3$ the lhs of \eqref{exp-Frs} vanishes and this implies
\be \label{div_U(k,D-k-3)}
(\Ddot)^{k-l} U^{(k,D-k+l-3)} = 0 \, ,
\ee
that for $l=0$ gives the constraint \eqref{eom_div} that we use in the evaluation of supertranslation charges.

\subsection{Stationary solutions}

In analogy with \cite{memory-anyD-Wald}, we identify stationary solutions of Fronsdal's equations as those satisfying
\be \label{stationary_def}
\pr_{u} U^{(k,n)} = 0 \quad \textrm{for} \quad n \leq D-k-3 \, , 
\ee
and we now prove that, up to pure-gauge contributions, this definition implies
\be \label{stationary}
U^{(k,n)} = 0 \quad \textrm{for} \quad n < D-k-3 \, .
\ee 
The latter condition corresponds to eq.~\eqref{synthCoul}, that we used in the evaluation of higher-spin supertranslations charges.

To begin with, we now prove that \eqref{stationary_def} implies $U^{(k,D-k-4)} = 0$. To this end, let us first consider \eqref{exp-Fus} for $k=0$ and $n=D-3$. Taking \eqref{stationary_def} into account, this gives
\be
0 = \left[ \Delta - D + 4 \right] U^{(0,D-4)} \qquad \Rightarrow \qquad U^{(0,D-4)} \, .
\ee
We can now proceed by induction on $k$. Assuming $U^{(l,D-l-4)}=0$ for $l<k$ one first obtains $\Ddot U^{(k,D-k-4)}$ from \eqref{exp-Frs} and, then, evaluating \eqref{exp-Fus} at $n=D-k-3$,
\be \label{U(k,D-k-4)}
0 = \left[ \Delta - (D+k-4) \right] U^{(k,D-k-4)} \qquad \Rightarrow \qquad U^{(k,D-k-4)} = 0 \, .
\ee
The last implication follows from \eqref{eigenvalues}, implying, in particular, that the eigenvalues of the Laplacian on an irreducible tensor of any rank are always negative. The same argument can be extended, modulo pure-gauge contributions, also to all $n < D-k-4$ with an induction in $n$. Assuming $U^{(0,n)} = 0$ eq.~\eqref{exp-Fus} gives indeed
\be
0 = \left[ \Delta - (n-1)(D-n-2) \right] U^{(0,n-1)} = \left[ \Delta + \ell(\ell + D-3 ) \right] U^{(0,D-3+\ell)} \, ,
\ee
which implies $U^{(0,D-3+\ell)} = 0$ for $\ell \leq -1$ or, equivalently, $U^{(0,n)} = 0$ for $n \leq D-4$. All tensors $U^{(k,n+k-1)}$ can then be set to zero with the same strategy that led to \eqref{U(k,D-k-4)}.

\subsection{Static solutions}

Static solutions are defined by
\be \label{static_app}
\pr_{u} U^{(k,n)} = 0 \quad \forall\ n \, , 
\ee
and we now argue that this condition implies, up to pure-gauge configurations,
\be \label{static_zero}
U^{(k,n)} = 0 \quad \textrm{for} \quad n < D-3
\ee 
together with
\be \label{killing_app}
\cK^{(k)} \equiv \cD\, U^{(k,D-3)} - \frac{2}{D+2(k-2)}\, \g\, \Ddot U^{(k,D-3)} = 0 \, .
\ee

For the $U^{(k,n)}$ with $1-k < n < D-k-3$ the same considerations as in the previous subsection apply, so that \eqref{stationary} is valid also for static solutions. Outside of this range, we have to go back to equations \eqref{exp-Frs} and \eqref{exp-Fus}. For $n = D-3$, they imply
\be \label{pre-killing}
0 = \left[ \Delta + (2k-1) (D-4) - k(D-k-2) \right] U^{(k,D-4)} - \cK^{(k-1)} \, ,
\ee
where $\cK^{(k)}$ denotes the conformal Killing equation for $U^{(k,D-3)}$ as in \eqref{killing_app}.
For $n = D-2$ one can use again \eqref{exp-Frs} for all tensors in the second line of \eqref{exp-Fus} to obtain
\be \label{div-killing}
0 = \Ddot \cK^{(k)} \, .
\ee

For static fields, the equations of motion thus set to zero the divergence of the conformal Killing tensor equation \eqref{killing} and on a compact space like the celestial sphere this suffices to set it to zero altogether. To support this statement let us notice that, assuming that the divergence of $U^{(k,D-3)}$ satisfies
\be \label{double-killing}
\left( \cD^2 + 2\,\g \right) \Ddot U^{(k,D-3)} - \frac{1}{D+2k-5}\,\g \left( \cD  - \frac{2}{D+2k-6}\, \g\, \Ddot \right) \Ddot\Ddot U^{(k,D-3)} = 0 \, ,   
\ee
the symmetrised gradient of \eqref{div-killing} can be rewritten as
\be \label{extra-identity}
\begin{split}
	0 = \cD \Ddot \cK^{(k)} & = 
	\left[ \Delta + (k-1)(k+D-4)-(k+1) \right]
	\cK^{(k)} + \frac{2}{D+2(k-2)}\, \g\, \Ddot\Ddot \cK^{(k)} .
\end{split}
\ee
The second term vanishes on shell, while the invertibility on $S^{D-2}$ of the differential operator in the first term guarantees 
\be \label{killing=0}
\cK^{(k)} = 0 \, .
\ee
In the previous step we used the identity \eqref{extra-identity} that one can prove with a similar approach starting from the double symmetrised gradient of $\Ddot\Ddot \cK^{(k)} = 0$ and progressing by recursion to identify a series of identities of the type $\cD^{k+1} (\Ddot)^k U^{(k,D-3)} = \g \left( \cdots \right)$. We do not have a complete proof that \eqref{div-killing} implies \eqref{killing=0}, but we checked that this is true up to $s=3$ and we shall assume this implication. 

Eq.~\eqref{pre-killing} then allows one to set to zero all $U^{(k,D-4)}$  and, via \eqref{exp-Frs}, also all $U^{(k,n)}$ in the range $D-k-4 \leq n \leq D-4$, thus providing the missing instances of \eqref{static_zero}.



\end{document}